\documentclass[journal]{IEEEtran}
\usepackage{graphicx}
\usepackage{color}
\usepackage{cite}
\usepackage{amsmath,amssymb}
\usepackage{threeparttable}

\usepackage{pgfplots}
\pgfplotsset{compat=1.9}
\usepackage{tabularx,booktabs}
\usepackage{wasysym}
\usepackage[bookmarks=false]{hyperref} %important for hypereference in the bibliography

\newcolumntype{C}{>{\centering\arraybackslash}X} % centered version of "X" type
\usepackage{siunitx}

\usepackage{tikz}
\usepackage{tikzscale}

\usepackage{dblfloatfix}
\usepackage{epstopdf}
\AppendGraphicsExtensions{.tif}

\usepackage{color}

\begin{document}

\title{A 3.5-THz, $\times 6$-Harmonic, Single-Ended Schottky Diode Mixer for Frequency Stabilization of Quantum-Cascade Lasers}

\author{Divya~Jayasankar,~\IEEEmembership{Graduate student~member,~IEEE,}
        Vladimir~Drakinskiy,
        Nick~Rothbart,
        Heiko~Richter, 
        Xiang~Lü,
        Lutz~Schrottke,
        Holger~T.~Grahn, 
        Martin~Wienold,
        Heinz-Wilhelm~Hübers,
        Peter~Sobis,
        and~Jan~Stake,~\IEEEmembership{Senior Member,~IEEE}% <-this % stops a space

\thanks{This research was carried out in the Gigahertz Center at Chalmers University of Technology in a project financed by VINNOVA, Chalmers University of Technology, Omnisys Instruments AB, Low  Noise Factory, Wasa Millimeter Wave, the Research Institutes of Sweden (RISE), and Virginia Diodes, Inc. The mixer development was financed in part by the European Space Agency (ESA) under the contract No. 4000125911/18/NL/AF "Frequency stabilisation of a Quantum Cascade Laser for Supra-THz applications" and in part by the Swedish National Space Board under the contract No. 170/17 "THz Schottky diode mixers for high-resolution FIR spectroscopy". \textit{(Corresponding author: Divya Jayasankar).} }
\thanks{D. Jayasankar, V. Drakinskiy, P. Sobis, and J. Stake are with the Terahertz and Millimetre Wave Laboratory, Department of Microtechnology and Nanoscience (MC2), Chalmers University of Technology, SE-412 96 Gothenburg, Sweden. D. Jayasankar is also with Research Institutes of Sweden (RISE), SE-504 62 Borås, Sweden. P. Sobis is also with Omnisys Instruments AB, SE-421 32 V\"{a}stra Fr\"{o}lunda, Sweden. (e-mail: divyaj@chalmers.se; vladimir.drakinskiy@chalmers.se; ps@omnisys.se; jan.stake@chalmers.se).}
\thanks{N. Rothbart, H. Richter, M. Wienold, and H.-W. Hübers are with the German Aerospace Center (DLR), Institute of Optical Sensor Systems, 12489 Berlin, Germany. (e-mail: nick.rothbart@dlr.de; heiko.richter@dlr.de; martin.wienold@dlr.de; heinz-wilhelm.huebers@dlr.de).}
\thanks{N. Rothbart, and H.-W. Hübers are also with the Humboldt-Universität zu Berlin, Department of Physics, 12489 Berlin, Germany.}
\thanks{X. Lü, L. Schrottke, and H. T. Grahn are with the Paul-Drude-Institut f\"ur Festk\"orperelektronik, Leibniz-Institut im Forschungsverbund Berlin e.~V., 10117 Berlin,  Germany (e-mail: lue@pdi-berlin.de; lutz@pdi-berlin.de; htgrahn@pdi-berlin.de).}}
%\thanks{Color versions of one or more of the figures in this article are available online at http://ieeexplore.ieee.org. Digital Object Identifier}}

\markboth{SUBMITTED TO IEEE TRANSACTIONS ON TERAHERTZ SCIENCE AND TECHNOLOGY, 2021}%
{Divya \MakeLowercase{\textit{et al.}}}

\maketitle

\begin{abstract}
Efficient and compact frequency converters are essential for frequency stabilization of terahertz sources. In this paper, we present a $\mathbf{3.5}$-THz, $\mathbf{\times 6}$-harmonic, integrated Schottky diode mixer operating at room temperature. The designed frequency converter is based on a single-ended, planar Schottky diode with a sub-micron anode contact area defined on a suspended 2-$\mathbf{\mu}$m ultra-thin GaAs substrate. The dc-grounded anode pad was combined with the radio frequency E-plane probe, which resulted in an electrically compact circuit. At $\mathbf{200}$ MHz intermediate frequency, a mixer conversion loss of about $\mathbf{59}$~dB is measured resulting in a $\mathbf{40}$~dB signal-to-noise ratio \textcolor{black}{for phase locking a $\mathbf{3.5}$-THz quantum-cascade laser}. Using a quasi-static diode model combined with electromagnetic simulations, good agreement with the measured results was obtained. Harmonic frequency converters without the need of cryogenic cooling will help in the realization of highly sensitive space and air-borne heterodyne receivers.
\end{abstract}


\IEEEpeerreviewmaketitle

\section{Introduction}

\IEEEPARstart{M}{olecular} emission spectroscopy is crucial for space and atmospheric sciences since it helps us to understand the stellar evolution, star formation, and cosmic chemistry \cite{Waters2006}. In particular, the observation of spectral signatures of molecular species such as the hydroxyl radical OH and atomic oxygen OI at terahertz (THz) frequencies ~\cite{Richter2021} provides valuable information about the Earth's atmosphere and global climate change. Future space and air-borne missions aiming at studying the chemical composition of the atmosphere in the far-infrared or THz spectral region will require reliable, high-resolution heterodyne receivers preferably operating at ambient temperatures \cite{Siegel2007},~\cite{Farrah2019}. 

In recent years, quantum-cascade lasers (QCLs) \cite{Koehler2002} have shown unprecedented improvement in performance, thereby making them ideal candidates for local oscillator (LO) sources for THz heterodyne receivers \cite{Richter2015},\textcolor{black}{~\cite{Kloosterman2013}}. However, QCLs are susceptible to frequency instabilities. Hence, it is of utmost importance to stabilize the signal from the QCL to a reference source to eradicate frequency jitters and to limit the phase noise. In 2005, Betz \textit{et al.} \cite{Betz} first demonstrated phase locking of a $3$-THz QCL with a far-infrared gas laser using the intermediate frequency (IF) signal generated by a GaAs Schottky diode mixer. In 2009, Rabanus \textit{et al.} \cite{Rabanus2009} reported phase locking of a $1.5$-THz QCL and the first heterodyne experiment using a phase-locked QCL as an LO source and a hot-electron bolometer (HEB) as a mixer, which requires cryogenic cooling for operation.

Frequency converters that can operate at ambient temperatures enable operation for a long lifetime and eliminate the necessity of bulky cryostats. \textcolor{black}{In 2009, Khosropanah \textit{et al.} demonstrated phase locking of a $2.7$-THz QCL using a superlattice mixer \cite{Khosropanah2009}}. Later in 2013, Hayton \textit{et al.} \cite{hayton2013} reported both, frequency and phase, locking of a $3.4$-THz QCL to a $\times 15$-harmonic signal generated by a superlattice harmonic mixer operating at room temperature. Subsequently, in 2014 they reported phase locking of a $4.7$-THz QCL to a superlattice harmonic mixer, which was cooled to 10~K and resulted in a beat signal with a signal-to-noise ratio (SNR) of 20~dB  \cite{khudchenko}.

Schottky diode-based harmonic mixers facilitate a broad IF range as well as a fast response time. Danylov \textit{et al.} \cite{Danylov2015} demonstrated phase locking of a $2.32$-THz QCL using a balanced-Schottky diode $\times 21$-harmonic mixer, which exhibited a conversion loss of about $110$~dB and an SNR of $25$~dB. Nonetheless, with increasing harmonic number, the SNR of the beat signal becomes degraded, thereby making it less suitable for applications that demand high sensitivity. Therefore, it is desirable to have THz harmonic mixers that exhibit a low conversion loss and can generate beat signals with a high SNR for QCL frequency stabilization. Bulcha \textit{et al.} \cite{bulcha2016} designed single-ended Schottky diode harmonic mixers yielding a conversion loss of 30~dB for fourth-harmonic mixing. 

Motivated by the performance of Schottky diodes at THz frequencies \cite{imran}, we have designed and developed a $3.5$-THz, $\times 6$-harmonic Schottky diode mixer for QCL frequency stabilization. The harmonic mixer design as well as the fabrication of mixer circuits and blocks were carried out at Chalmers University of Technology, the QCL was fabricated at Paul-Drude-Institut (PDI), and the mixer characterization was performed at the German Aerospace Center (DLR). The article is organized as follows: the design and development of the $\times 6$-harmonic, single-ended Schottky diode mixer are presented in section II. The mixer characterization setup is described in section III. Finally, results from the dc measurements and radio frequency (RF) characterization are presented and discussed in section IV.

\section{Method}
\label{sectionII}

\begin{figure}[!b]
\centering    
\includegraphics[width=0.5\textwidth,keepaspectratio]{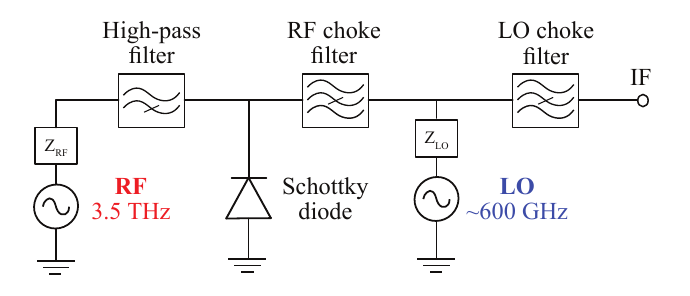}
    \caption{Harmonic-mixer topology. Equivalent circuit of the $3.5$-THz, $\times$6-harmonic, planar, single-ended Schottky diode mixer. $Z_{\text{RF}}$ and $Z_{\text{LO}}$ are the embedding impedances of the diode at the RF and LO frequency, respectively. Optimization of the circuit elements and waveguide split blocks was carried out to present appropriate embedding impedances to the diode to obtain low mixer conversion loss.}
    \label{figure1}
\end{figure}

This section describes the diode modeling, the $\times$6-harmonic mixer design, the integrated-diode fabrication process, the mixer housing fabrication, and, finally, the process of the circuit assembly. First, for the chosen mixer topology, the diode modeling was carried out to evaluate the diode dc parameters. Second, a large-signal, harmonic-balance simulation in the circuit analyzer was set up to determine the diode-embedding impedances at the RF, LO frequency, and IF. Thereafter, a three-dimensional (3D) electromagnetic (EM) model of the waveguide and filter sections was designed using a finite-element method (FEM) solver. Upon optimization, the overall performance of the harmonic mixer was analyzed in terms of the RF and LO return losses as well as the mixer conversion loss \textit{L}. Later, the designed mixer circuit was realized on an ultra-thin, semi-insulating GaAs substrate. Simultaneously, the mixer blocks were manufactured using a high-speed micrometer precision milling tool. Finally, the integrated mixer circuit on the GaAs substrate, the quartz carrier substrate, and the printed circuit board (PCB) were assembled on the E-plane mixer split block.

\subsection{Diode modeling}
\label{sectionIIA}
At $3.5$~THz, the estimated losses in the RF rectangular waveguide WM-64\footnote{RF waveguide name designation. 'W' stands for waveguide, 'M' for metric, and the number is the waveguide width in $\mu$m \cite{Waveguide}.} are expected to be higher than $1$~dB/mm. Hence, it is of utmost importance to find a compact solution that takes into account the high-frequency losses and the narrow-tolerance limit set by the fabrication process. Therefore, a single-ended topology was chosen to realize the $3.5$-THz, $\times$6-harmonic mixer as illustrated in Fig.~\ref{figure1}. 

\begin{figure}[!t]
\centering
    \includegraphics[width=0.5\textwidth,keepaspectratio]{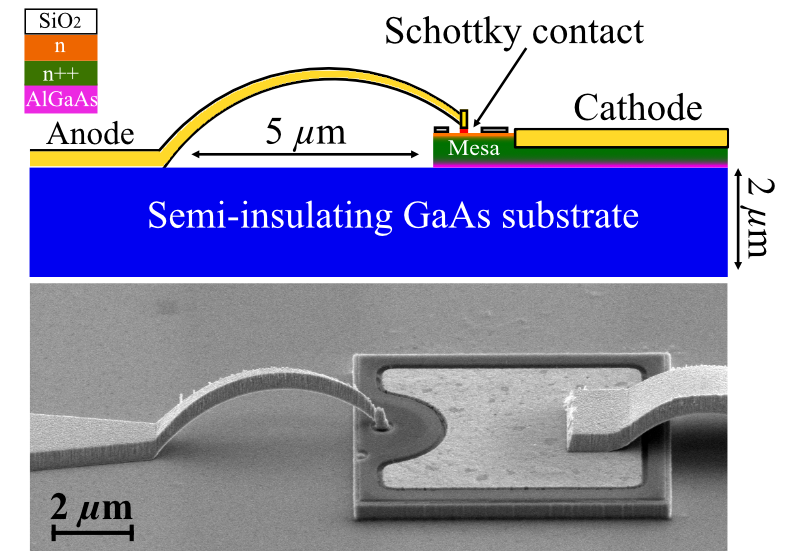}
    \caption{Single-ended Schottky diode contact. Top: Illustration of the Schottky contact with the sub-micron anode area (Note: The drawing is not to scale). Bottom: Scanning electron micrograph of a planar air-bridge Schottky diode with a contact area $S = 0.11~\mu$m$^2$. }
    \label{figure2} 
\end{figure}

A planar, air-bridged Schottky diode with a sub-micron anode area was defined on a semi-insulating GaAs substrate as shown in Fig.~\ref{figure2}. A $50$-nm-thick epilayer with a doping concentration of about $6 \times 10^{17}$~cm$^{-3}$ was chosen in order to operate below the plasma frequency \cite{crowe1992}. For the initial mixer design, a quasi-static, equivalent-lumped-circuit diode model that consists of the diode series resistance ($R_s$), the zero-bias junction capacitance ($C_{j0}$), and a parallel non-linear current source is applied to the single-ended discrete Schottky diode. For a Schottky contact with an area of 0.11~$\mu$m$^2$, a series resistance of about 25~$\Omega$ was calculated using the analytical model which takes into account the undepleted epilayer resistance \cite{crowe1992}, dc spreading resistance in the buffer layer \cite{dickens}, and ohmic contact resistance. The effect of self-heating is not included \cite{Tang2012}. In the junction capacitance model, the first-order fringing effect\textcolor{black}{\cite{Diego2016}} was included as shown below\cite{louhi1994} , 

\begin{equation}
    C_j \simeq \frac{\epsilon_l S}{W_{d\text{0}}} \bigg( 1+ 2.5 \bigg( \frac{W_{d\text{0}}}{\sqrt{S}} \bigg) \bigg),
    \label{equ1}
\end{equation}

\noindent where $\epsilon_l$ denotes the relative permittivity of the $n$-doped GaAs layer and $S$ the \textcolor{black}{rectangular area of the Schottky contact} . The depletion width $W_{d\text{0}}$ at zero bias is assumed to be equal to the thickness of the \textit{n}-doped semiconductor layer. In the ideal-diode model, an ideality factor $\eta = 1.2$, a saturation current $I_{\text{sat}} = \SI{1}{fA}$, and a built-in potential of \SI{0.85}{V} were assumed.

\subsection{Design of the $\times$6-harmonic mixer}

\begin{figure}
    \centering
    \includegraphics[width= 0.75\linewidth]{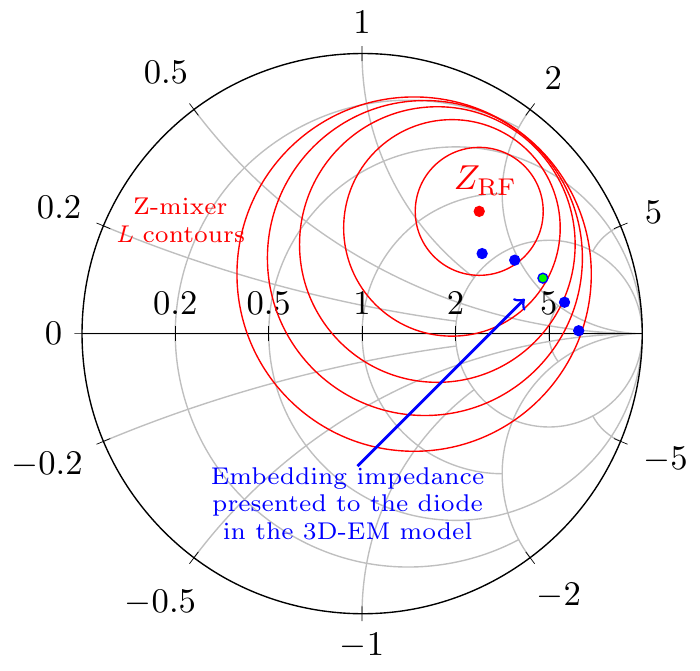}
    \caption{RF embedding impedance. The Smith chart shows the optimum RF embedding impedance $Z_{\text{RF}}$ indicated by a red dot and the conversion loss contours (1~dB step) derived from the Z-mixer simulation. The RF embedding impedance presented to the diode in the EM simulation software in the frequency range from 3.4 to 3.6~THz is indicated by blue dots and the center frequency is highlighted by a green dot.}
    %that yielded a conversion loss of 24~dB excluding the 3D-EM model. 
    \label{fig:smith}
\end{figure}

The embedding impedances of a Schottky diode with a sub-micron anode area were evaluated using a large-signal, harmonic-balance simulation. Using the built-in optimizer in the circuit simulator, the diode-embedding impedances at the RF, IF, and LO frequency were varied to provide a low mixer conversion loss \textit{L}. The out-of-band frequencies were terminated by an open circuit (Z-mixer topology) \cite{Saleh},~\cite{Divya2019}. The RF optimum diode-embedding impedance of the Z-mixer is approximately $Z_{\text{RF}} = (60+j80)~\Omega$ as shown in Fig.~\ref{fig:smith}. \textcolor{black}{At the LO and IF the optimum impedances are approximately $Z_{\text{LO}} = (150+j300)~\Omega$ and 1000~$\Omega$ respectively \cite{Divya2019}}. \textcolor{black}{The predicted conversion loss $L$ from the ideal Z-mixer circuit simulation was about 20~dB for a Schottky contact area of 0.11~$\mu$m$^2$, an LO power = 2~dBm, no dc bias, and the losses from the mixer circuit were excluded.}

\begin{figure}[!b]
\centering
    \includegraphics[width=0.5\textwidth]{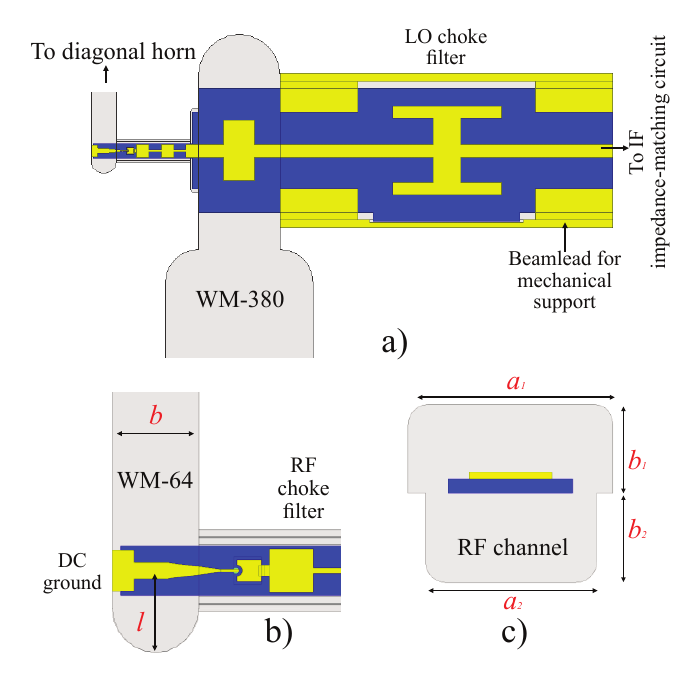}
    \caption{EM model for the 3.5-THz, $\times$6-harmonic Schottky diode mixer. (a) Full 3D-EM model of the integrated planar Schottky diode realized on a $2$-$\mu$m-thick GaAs substrate, which is suspended on an E-plane split block using beam leads. (b) RF chain consisting of a WM-64 waveguide, dc grounded RF E-plane probe, Schottky diode contact, and RF choke filter. (c) Cross section image of the RF filter channel. }
    \label{figure3}
\end{figure} 
Based on the topology illustrated in Fig.~\ref{figure1}, a 3D-EM model of the choke filters and matching networks was implemented using a FEM solver (Ansys HFSS) as shown in Fig.~\ref{figure3}(a). The RF waveguide (WM-64), LO waveguide (WM-380) and \textcolor{black}{IF port} were assigned with a waveguide port. A $50$-$\Omega$ lumped port was defined at the Schottky diode junction. To evaluate the circuit performance accurately, multi-frequency adaptive meshing was carried out at the RF and LO frequency. When the solver attains the specified convergence criteria, it continues to refine the mesh for five consecutive adaptive passes, which resulted in a total of about $200\,000$ tetrahedrons. To reduce the computation time, the diagonal horn and the WM-380 LO access waveguide were excluded. 
A reduced gold conductivity of about $2 \times 10^{7}$~S/m was assigned to the metal strip lines and waveguide walls \cite{Laman2008}. The GaAs substrate with a relative permittivity $\epsilon_s = 12.9$ and a loss tangent $\tan~\delta = 0.001$ was used in the EM-simulations \cite{Palik1997},~\cite{Grischkowsky}. Additional parasitic elements of the diode were included in the 3D-EM model, including high-frequency losses in the contact mesa due to a limited conductivity of the GaAs buffer layer of about $\sigma = q~\mu~N_{\text{buffer}}$ = \SI{1.5e5}{S/m}. Note: A surface impedance boundary condition was applied to the mesa so fields are not solved inside.

The incoming THz signal from the QCL is coupled to the diode using an RF E-plane probe. \textcolor{black}{For maximum RF energy coupling efficiency to the diode and to present the optimum embedding impedance at the RF}, a diode geometry optimization and an RF backshort tuning were carried out as shown in Fig.~\ref{figure3}(b). Thereafter, an RF choke filter was implemented as a high-low impedance line to prevent the leakage of the RF signal into the LO chain. Fig.~\ref{figure3}(c) shows the cross section of the RF channel with a $2$-$\mu$m-thick GaAs substrate and a $0.5$-$\mu$m-thick gold metalization layer. The corner radii in the RF channel arising from the milling process were taken into account.

\begin{figure}[!t]
    \centering
    \includegraphics[width=0.95\linewidth]{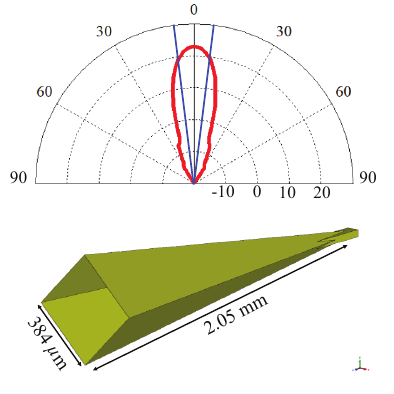}
    \caption{\textcolor{black}{3.5-THz diagonal horn. Top: A polar plot showing the simulated  radiation pattern (H-plane) in red and blue lines shows the 3-dB angular width. Bottom: Layout of the diagonal horn antenna used for the 3D-EM model.
}}
    \label{fig:horn}
\end{figure}

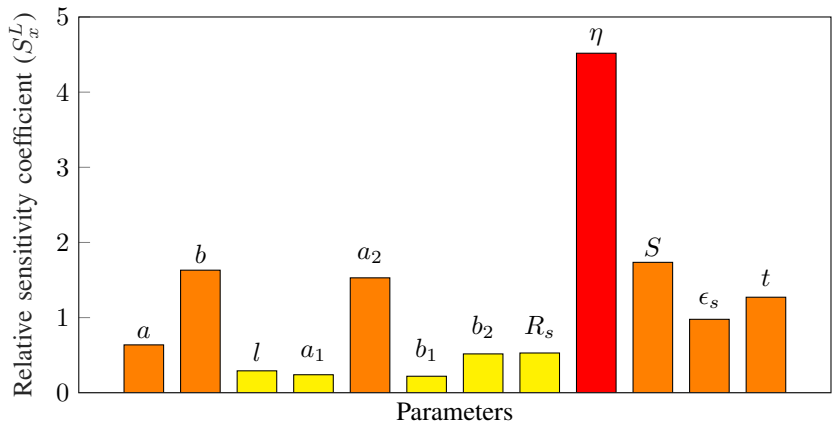
\begin{figure*}[!t]
\centering
\begin{tikzpicture}

\begin{axis}[%
width=10cm,
height=5cm,
at={(0.758in,0.481in)},
scale only axis,
bar shift auto,
xmin=-0.15,
xmax=13.15,
xtick={-0.2 14},
ytick ={0,1,2,3,4,5},
ymin=0,
ymax=5,
xtick pos = left, 
ytick pos = left,
xtick align = inside,
ytick align = inside,
bar shift = 0 pt,
ylabel style={font=\color{white!15!black}},
ylabel={Relative sensitivity coefficient ($S_x^L$)},
xlabel={Parameters},
axis background/.style={fill=white},
legend style={legend cell align=left, align=left, draw=white!15!black}
]
\addplot[ybar, bar width=0.7, draw=black, area legend, fill = orange] table[row sep=crcr] {%
1	0.637244180086383\\
};
\addplot[ybar, bar width=0.7, draw=black, area legend, fill = orange] table[row sep=crcr] {%
2	1.63117008273886\\
};
\addplot[ybar, bar width=0.7, draw=black, area legend, fill = yellow] table[row sep=crcr] {%
3	0.292005271944286\\
};
\addplot[ybar, bar width=0.7, draw=black, area legend, fill = yellow] table[row sep=crcr] {%
4	0.240001022367003\\
};
\addplot[ybar, bar width=0.7, draw=black, area legend, fill =orange] table[row sep=crcr] {%
5	1.52922195256768\\
};
\addplot[ybar, bar width=0.7, draw=black, area legend, fill = yellow] table[row sep=crcr] {%
6	0.219995562530921\\
};
\addplot[ybar, bar width=0.7, draw=black, area legend, fill = yellow] table[row sep=crcr] {%
7	0.517196785351497\\
};
\addplot[ybar, bar width=0.7, draw=black, area legend, fill = yellow] table[row sep=crcr] {%
8	0.529312128434398\\
};
\addplot[ybar, bar width=0.7, draw=black, area legend, fill = red] table[row sep=crcr] {%
9	4.51777295023126\\
};
\addplot[ybar, bar width=0.7, draw=black, area legend, fill = orange] table[row sep=crcr] {%
10	1.7354568249129\\
};
\addplot[ybar, bar width=0.7, draw=black, area legend, fill = orange] table[row sep=crcr] {%
11	0.977412902737142\\
};
\addplot[ybar, bar width=0.7, draw=black, area legend, fill = orange] table[row sep=crcr] {%
12	1.27135966485938\\
};

\addplot[forget plot, color=white!15!black] table[row sep=crcr] {%
-0.15	0\\
13.15	0\\
};
 \node [above] at (axis cs: 1,0.6) {$a$};
 \node [above] at (axis cs: 2,1.6) {$b$};
 \node [above] at (axis cs: 3,0.3) {$l$};
 \node [above] at (axis cs: 4,0.3) {$a_{1}$};
 \node [above] at (axis cs: 5,1.6) {$a_{2}$};
 \node [above] at (axis cs: 6,0.3) {$b_1$};
 \node [above] at (axis cs: 7,0.6) {$b_2$};
 \node [above] at (axis cs: 8,0.6) {$R_s$};
 \node [above] at (axis cs: 9,4.5) {$\eta$};
 \node [above] at (axis cs: 10,1.7) {$S$};
 \node [above] at (axis cs: 11,1) {$\epsilon_s$};
 \node [above] at (axis cs: 12,1.3) {$t$};

\end{axis}
\end{tikzpicture}%
    \caption{Robustness of the 3.5-THz, $\times$6-harmonic Schottky diode mixer. A sensitivity analysis showing the influence of the circuit design and diode model parameters listed in Table~\ref{table1}. The following parameters were assigned to the harmonic-balance simulation: RF = 3.5~THz, IF = 5~GHz, LO power = 2 dBm, RF power = \SI{-50}{dBm}, and no dc bias.}
    \label{figure4}
\end{figure*}

%%%%%%%%%%%%%%%%%%%%%%%%%%%%%%%%%%%%%%%%%%%%%%%%%%%%%%%%%%%%%%%%%%%%%%%%%%%%%%%%%%%%
\begin{table*}[!b]
\center
\begin{threeparttable}[b]
\caption{List of the Circuit Parameters Studied in the Sensitivity Analysis}
\label{table1}
\centering
\begin{tabularx}{\textwidth}{@{}l*{10}{C}c@{}}

\toprule
 & RF waveguide width & $a$ = 2$b$ & $64$ $\mu$m   \\
 & RF waveguide height & $b$ & $32$ $\mu$m   \\
  & RF backshort & \textit{l} & $30$ $\mu$m   \\
 Block & RF filter channel width (top) & $a_1$ & $30$ $\mu$m  \\
& RF filter channel width (bottom) & $a_2$ & $25$ $\mu$m   \\
 & RF filter channel height (top) & $b_1$ & $10$ $\mu$m  \\
  & RF filter channel height (bottom) & $b_2$ & $10$ $\mu$m  \\
\midrule
 & Series resistance & $R_s$ & $25$ $\Omega$   \\
Diode & Ideality factor & $\eta$ & 1.2   \\
 & Schottky junction area & $S$ & 0.11~$\mu$m$^2$   \\
\midrule
GaAs substrate & Relative permittivity & $\epsilon_s$ & 12.9~~\cite{Palik1997} \\
& Substrate thickness & \textit{t} & $2$ $\mu$m  \\
\bottomrule
\end{tabularx}
%\begin{tablenotes}[flushleft]
%\end{tablenotes}
\end{threeparttable}
\end{table*}

The same procedure as described earlier was followed to present the LO optimum embedding impedance to the diode. Thereupon, a hammer head filter that prevents the LO signal to propagate along the IF line was designed. Both, RF and LO, channel dimensions were carefully varied such that it only allows the fundamental, quasi-transverse electromagnetic mode to propagate along the \textcolor{black}{planar circuit}. In addition, the asymmetrical GaAs substrate and the alignment pockets in the waveguide blocks facilitate a precise alignment of the GaAs membrane during the circuit assembly process. Upon optimization of the EM model, the four-port \textit{S}-parameters were imported to the circuit simulator, where the model was analyzed using a large-signal, harmonic-balance simulation with a standard diode model. Finally, the overall mixer performance was evaluated by taking into account the metal losses in the stripline and waveguide. Based on the ideal-diode model described in Section~\ref{sectionIIA} and including the 3D-EM model with parasitic effects as described above, we predict a conversion loss of about 45~dB at 3.5~THz.

The RF feedhorn is based on a standard diagonal horn design described by Johansson and Whyborn \cite{horn} and was verified using an FEM solver. A diagonal horn with an aperture size of $384$$\times$$384~\mu$m$^2$, corresponding to a flare angle of $5.4^\circ$ and length 2.05 mm as shown in Fig.~\ref{fig:horn}, resulted in a simulated directivity of about $23$~dBi. Finally, a three-section Chebyshev impedance transformer was designed to \textcolor{black}{transform the 50-$\Omega$ IF output to $150~\Omega$, which is close to the practical limit of a microstrip line on a Rogers 4003B printed circuit board.} The simulated insertion loss was less than $0.3$~dB in the frequency range from 1.5 to 7~GHz.

\subsection{Sensitivity analysis} \label{sectionIIC}
To check the robustness of the design, a sensitivity analysis was performed. The relative sensitivity coefficient \textit{$S_x^L$} is defined as the ratio of the relative change in the output to the relative change in the input variable \cite{Smith2007}: 

\begin{equation}
    S_x^L = \frac{\Delta L/L_0}{\Delta x/x_0}
    \label{equation}
\end{equation}
where $\Delta L = L_x - L_0$ denotes the change in the conversion loss, $L_0$ the nominal conversion loss, $\Delta x$ the relative variation of the parameter in study, and $x_0$ the nominal value of the parameter. 

In this analysis, the design parameters with narrow-tolerance limits and with some influence on the mixer conversion loss were taken into account and are summarized in Table~\ref{table1}. The selected variables were then increased by 10\% of their initial values. The corresponding change in the mixer conversion loss at 3.5~THz was analyzed and is shown in Fig.~\ref{figure4}. \textcolor{black}{Changing the anode area \textit{S} will scale the dc current and also the junction capacitance $C_j$}. Among the critical parameters listed above, an increase of the ideality factor $\eta$ results in a weaker non-linearity and hence a noticeable difference in the mixer conversion loss.

\subsection{Integrated-diode fabrication process} 
The integrated mixer circuit \cite{siegel1999} was realized on a GaAs wafer that comprises a 650-$\mu$m-thick, semi-insulating, 3-inch GaAs substrate supporting a 2-$\mu$m-thick, semi-insulating GaAs membrane layer sandwiched between two (Al,Ga)As etch stop layers. The top (Al,Ga)As etch stop layer is followed by a $500$-nm-thick, heavily doped \textit{n}$^{++}$ buffer layer with a doping concentration of \SI{5e18}{cm^{-3}} and a $50$-nm-thick $n$-doped active layer with a doping concentration of \SI{6e17}{cm^{-3}}. All layers are grown by molecular beam epitaxy. 

 \begin{figure}[!b]
\centering
    \includegraphics[width=0.45\textwidth,keepaspectratio]{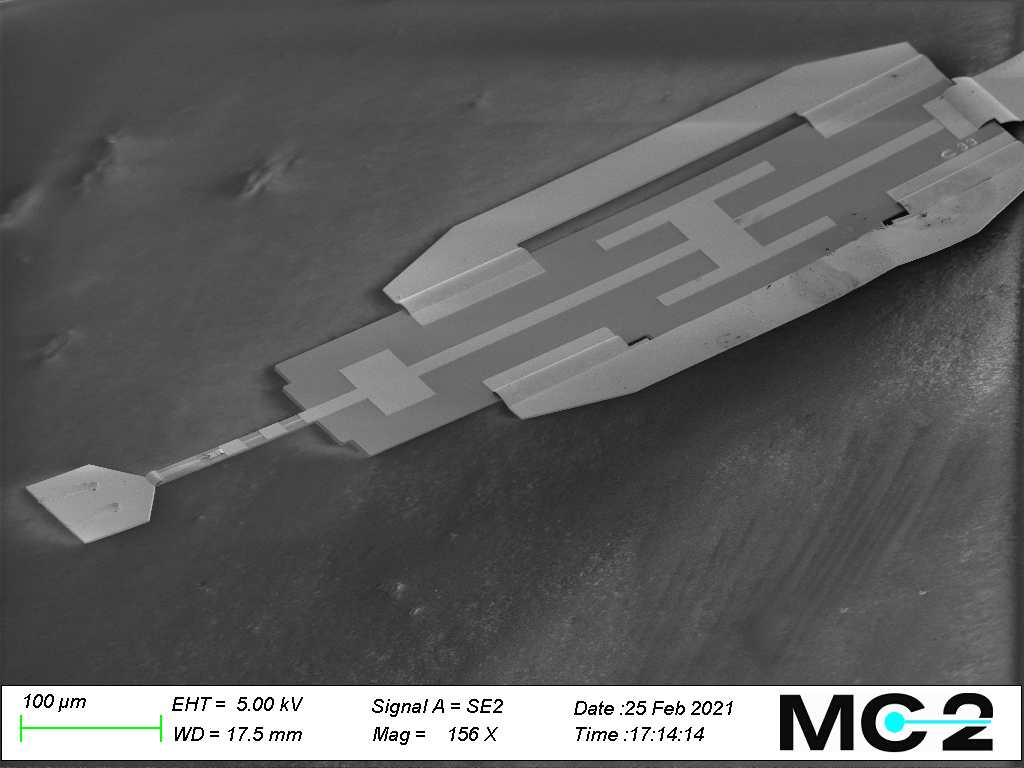}
    \caption{Integrated mixer circuit. Scanning electron micrograph of the 3.5-THz, $\times$6-harmonic Schottky diode mixer circuit after releasing it from the supporting GaAs substrate.}
    \label{figure5}
\end{figure}

Electron beam lithography was used to define each pattern. Firstly, a thin layer of SiO$_2$ is deposited on the membrane, which acts as a protective mask for the active layer during the initial process steps. Secondly, the ohmic and Schottky contacts were formed by deposition of Pd/Ge/Au/Pd/Au and Ti/Pt/Au metal layers, respectively. Next, the mesa was formed by dry etching in a silane based, inductively coupled plasma (ICP). Thereafter, the air bridge connections are formed by evaporation of a $500$-nm-thick gold layer. In the next step, diodes are isolated by a combination of selective and non-selective wet etching through the top (Al,Ga)As etch stop layer and the GaAs membrane down to the bottom (Al,Ga)As etch stop layer. Subsequently, the passive circuitry (beam leads, waveguide probes, and filter structures) are formed by evaporation of a gold film and followed by a lift-off process. Fig.~\ref{figure5} shows the integrated-diode mixer circuitry fabricated at Chalmers University of Technology. Integrated mixer circuits with three different Schottky contact areas (0.11, 0.14, and 0.17 $\mu$m$^2$) in rectangular shape were fabricated.

\begin{figure}[!b]
\centering
    \includegraphics[width=0.45\textwidth]{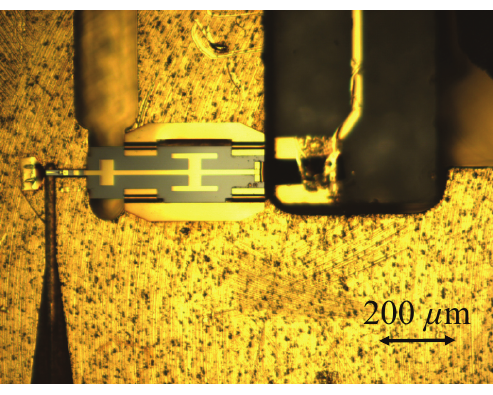}
    \caption{Mixer assembly. Optical micrograph of the assembled 3.5-THz, $\times$6-harmonic Schottky mixer showing the RF circuitry and the beam lead punched into the mixer housing that provides the dc ground. } %clamped to the mixer block to provide
    \label{assembly}
\end{figure}

\vspace{-0.2cm}

\subsection{Mixer housing fabrication}
The mixer was machined in an E-plane split aluminum block using a high-speed, micrometer precision CNC milling tool (KERN Evo). Four guide structures were defined on the top surface of the split block to ensure the precise alignment of the blocks during the assembly process. Care was taken to ensure that the milling tool was well aligned in the spindle using a dial indicator, and the alignment of the block to the spindle axis was calibrated. The sidewall of the mixer block that consists of a diagonal horn aperture was slanted by a $10^\circ$ angle to redirect the reflected incoming signal. 

First, the large features such as the PCB cavity, the quartz carrier cavity, as well as the LO waveguide and channel were machined to avoid the risk of leaving the machined chips inside the smaller features. Then, the RF waveguide and RF channel were machined using a 20-$\mu$m end mill tool that has a tolerance of $\pm$ 2 $\mu$m. Thereafter, the RF diagonal feedhorn was machined using a $45^\circ$ chamfering milling tool of $100~\mu$m diameter. To avoid unnecessary reflections caused due to burrs along the edge of machined features, the mixer block was cleaned with an acetone and iso-propanol solution in an ultrasonic bath, etched in an aluminum bath, and, finally, sputtered with a $0.5$-$\mu$m-thick gold layer as shown in Fig.~\ref{assembly}.

\subsection{Circuit assembly}
\begin{figure}[!t]
    \centering
    \includegraphics[width=0.45\textwidth]{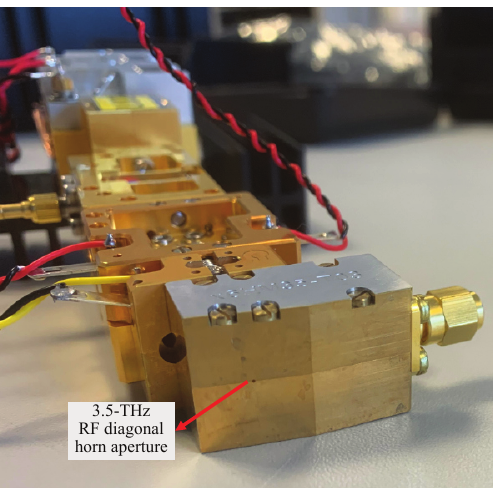}
    \caption{Mixer and LO multiplier chain assembly. Photograph of the assembled 3.5-THz, $\times$6-harmonic mixer and connected with a $600$-GHz, $\times 64$-active-frequency multiplier chain.}
    \label{fig:x6hm35}
\end{figure}

The quartz carrier substrate and PCB were mounted using a thin layer of conductive silver epoxy glue (Epotec-H20E), which was followed by soldering the IF coaxial connector pin and wire bonding the substrate interfaces. In the next assembly step, the membrane circuit was mounted in the E-plane of the mixer split block and precisely aligned with the aid of a high-magnification (500$\times$) assembly microscope. Thereafter, the RF, dc ground, and the IF beam leads were stamped to the mechanical housing and the IF pad on the quartz carrier substrate, respectively, using a standard ceramic wedge and ultrasonic compression bonding. Finally,\textcolor{black}{the top and bottom half split blocks were aligned using mechanically integrated alignment guide structures in the housing} and tightened with screws as shown in Fig.~\ref{fig:x6hm35}. 

\begin{figure} [!t]
\centering
    \includegraphics[width=0.5\textwidth]{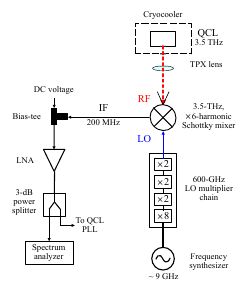}
    \caption{Illustration of the mixer characterization and QCL phase-locking measurement setup.}
    \label{setup}
    \end{figure}

     \begin{figure} [!t]
\centering
    \includegraphics[width=0.5\textwidth]{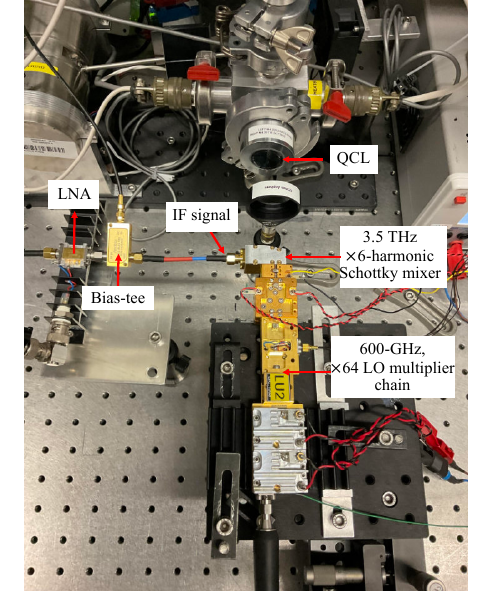}
    \caption{Harmonic mixer characterisation setup. Photograph of the 3.5-THz, $\times6$-harmonic mixer characterization setup at DLR, Berlin, showing the 3.5-THz QCL in a cryocooler, $\times$6-harmonic mixer, and 600-GHz LO multiplier chain. A dc voltage was applied via a bias-tee.}
    \label{setup1}
\end{figure} 

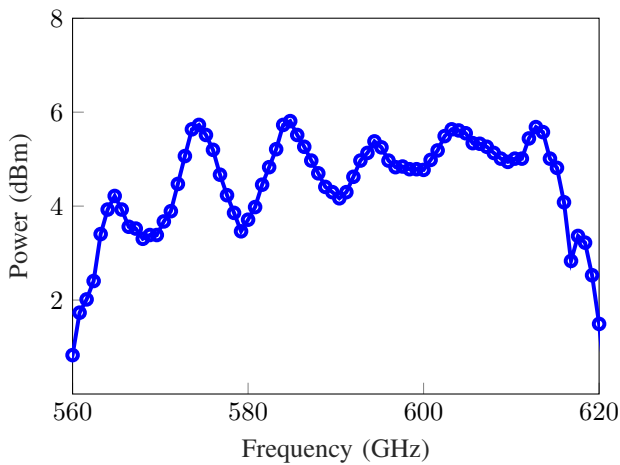
\begin{figure} [!b]
\centering
\definecolor{mycolor1}{rgb}{0,0,1}%
\begin{tikzpicture}

\begin{axis}[%
width=7cm,
height=5cm,
at={(0.758in,0.481in)},
scale only axis,
unbounded coords=jump,
xmin=560,
xmax=620,
xlabel style={font=\color{white!15!black}},
xlabel={Frequency (GHz)},
ymin=0,
ymax=8,
xtick pos = left, 
ytick pos = left,
xtick align = inside,
ytick align = inside,
ytick ={2,4,6,8},
xtick = {560,580,600,620},
ylabel style={font=\color{white!15!black}},
ylabel={Power (dBm)},
axis background/.style={fill=white},
legend style={legend cell align=left, align=left, draw=white!15!black}
]
\addplot [color=mycolor1, line width=1.5pt, mark=o, mark size=2pt, mark options={solid, mycolor1}]
  table[row sep=crcr]{%
552	2.09515014542631\\
552.8	1.98657086954423\\
553.6	1.64352855784437\\
554.4	1.39879086401236\\
555.2	1.98657086954423\\
556	0.863598306747482\\
556.8	0.293837776852097\\
557.6	0.530784434834197\\
558.4	0.211892990699381\\
559.2	0.334237554869497\\
560	0.827853703164501\\
560.8	1.73186268412274\\
561.6	2.01397124320451\\
562.4	2.405492482826\\
563.2	3.40444114840118\\
564	3.92696953259666\\
564.8	4.21603926869831\\
565.6	3.92696953259666\\
566.4	3.56025857193123\\
567.2	3.52182518111362\\
568	3.30413773349191\\
568.8	3.38456493604605\\
569.6	3.38456493604605\\
570.4	3.67355921026019\\
571.2	3.89166084364533\\
572	4.47158031342219\\
572.8	5.06505032404872\\
573.6	5.63481085394411\\
574.4	5.7287160220048\\
575.2	5.51449997972875\\
576	5.19827993775719\\
576.8	4.66867620354109\\
577.6	4.23245873936808\\
578.4	3.85606273598312\\
579.2	3.46352974450639\\
580	3.71067862271736\\
580.8	3.97940008672038\\
581.6	4.45604203273598\\
582.4	4.82873583608754\\
583.2	5.21138083704036\\
584	5.7287160220048\\
584.8	5.80924975675619\\
585.6	5.51449997972875\\
586.4	5.26339277389844\\
587.2	4.96929648073215\\
588	4.69822015978163\\
588.8	4.40909082065218\\
589.6	4.29752280002408\\
590.4	4.16640507338281\\
591.2	4.29752280002408\\
592	4.62397997898956\\
592.8	4.96929648073215\\
593.6	5.13217600067939\\
594.4	5.37819095073274\\
595.2	5.25044807036845\\
596	4.96929648073215\\
596.8	4.82873583608754\\
597.6	4.84299839346786\\
598.4	4.78566495593843\\
599.2	4.78566495593843\\
600	4.77121254719662\\
600.8	4.98310553789601\\
601.6	5.18513939877887\\
602.4	5.49003262025788\\
603.2	5.63481085394411\\
604	5.61101383649056\\
604.8	5.55094448578319\\
605.6	5.34026106056135\\
606.4	5.32754378992498\\
607.2	5.26339277389844\\
608	5.13217600067939\\
608.8	5.01059262217751\\
609.6	4.94154594018443\\
610.4	5.01059262217751\\
611.2	5.01059262217751\\
612	5.44068044350276\\
612.8	5.68201724066995\\
613.6	5.57507201905658\\
614.4	5.01059262217751\\
615.2	4.81442628502305\\
616	4.0823996531185\\
616.8	2.8330122870355\\
617.6	3.3645973384853\\
618.4	3.22219294733919\\
619.2	2.52853030979893\\
620	1.4921911265538\\
620.8	0\\
621.6	-2.00659450546418\\
};

%\addplot [only marks, mark=asterisk, mark size=3pt, mark options={red}]
  %table[row sep=crcr]{%
  %573.8473032012388 3.548173534904226 \\
%574.1257525123247 3.5367797271626666 \\
%574.5788893629124 3.4275560024920098 \\
%575.4063566552901 3.304565140840924 \\
%576.2338239476678 3.0019778999050777 \\
%577.0612912400455 2.6165000223762647 \\
%592.2709281380356 2.6998409813052193 \\
%595.6596037163443 2.698999074329242 \\
%};

\end{axis}
\end{tikzpicture}%
    \caption{Power versus frequency. Measured output power from the $600$-GHz, $\times$64-active frequency multiplier chain. At $573$~GHz, the measured output power is approximately \SI{5.6}{dBm}.}
    \label{lomulmeasurement}
\end{figure}

\section{Measurement setup}

The schematic of the harmonic mixer characterization measurement setup is shown in Fig.~\ref{setup}. The QCL developed and fabricated at Paul-Drude-Institut is based on a GaAs/AlAs heterostructure \cite{Schrottke2016}. This materials system allows for THz QCLs with a relatively high wall-plug efficiency and reduced cooling power so that operation in a compact Stirling cooler becomes feasible. For the active region, a hybrid design, which is preferred for continuous-wave operation, is employed. The design has been optimized for emission at $3.5$~THz with a sufficient frequency tuning range. The essential parameters are given in \cite{Schrottke2020}. The resonator is a Fabry-Pérot cavity based on a single-plasmon waveguide with a length of $826~\mu$m, a width of $120~\mu$m, and a height of $10~\mu$m. The QCL is placed in a compact Stirling cooler (AIM SL400) \textcolor{black}{which facilitates thermal stabilization using an internal temperature control for the cold finger inside the cryocooler and a Cernox temperature sensor mounted closely to the QCL. Typical operating temperature of the QCL is around 50 K with temperature stability of about $\pm$ 1 mK.}

%The incoming signal from the QCL was varied using a rotatable and a fixed polarizer, which allows for a controlled relative attenuation of the input RF power.

A TPX lens was used to focus the incoming THz signal from the QCL onto the integrated diagonal horn of the harmonic mixer. The mixer is pumped by the LO signal generated from the $600$-GHz, $\times$64-cascaded active multiplier chain. It consists of a $\times$8 E-band active multiplier (AMC-12-RNHB1) from Millitech (Smiths Interconnect), followed by a high-power isolator (HMI12-387-69.5-5.0) from HXI, and a cascaded three-stage frequency multiplier chain \textcolor{black}{design from Omnisys Instruments based on Chalmers GaAs Schottky membrane} diode varactor doublers for monolithic microwave integrated circuits. The available output power of the $600$-GHz, $\times$64 active frequency multiplier chain at the waveguide interface was measured using an Erickson power meter (PM5). A waveguide taper transition was used to connect the LO multiplier chain and the power meter, which have WR-1.5\footnote{WR - Rectangular waveguide, the number is the waveguide width in mils multiplied by 10. (1 mils = 1/1000 inch).} and WR-10 waveguide interfaces, respectively. The measured output power from the multiplier chain is corrected for conductor losses in this transition and plotted versus frequency as shown in Fig.~\ref{lomulmeasurement}. The output power generated by the 600-GHz multiplier chain can be controlled by either detuning the bias of the multiplier stages or by reducing the input power from the frequency synthesizer.

The $6^{th}$ harmonic of the LO signal at \SI{3.4424}{THz} is combined with the QCL radiation at \SI{3.4426}{THz} to generate an IF signal at \SI{200}{MHz}. A dc voltage is applied to the mixer via a bias-tee connected to an SMA connector to allow for optimization of the IF signal power. The generated output IF signal is amplified using a low-noise amplifier operating at room temperature (Miteq AFS4 00100600-1310P-4). Finally, the amplified IF signal is detected using a spectrum analyzer. Using a power splitter, part of the signal can be used for frequency stabilization of the QCL.

\section{Results and Discussion}

Two integrated circuits with Schottky contact areas of 0.11 and 0.14 $\mu$m$^2$ were assembled on mixer modules and characterized, showing similar performance \cite{DivyaIRMMW}. In this section, we will present the results of the mixer with a Schottky contact area of 0.11 $\mu$m$^2$. The experimental results are then compared to large-signal simulations, and finally, a summary of different state-of-the-art mixers available for the QCL frequency stabilization is presented.

\subsection{dc measurements}

\begin{figure}[h]
\centering
\input{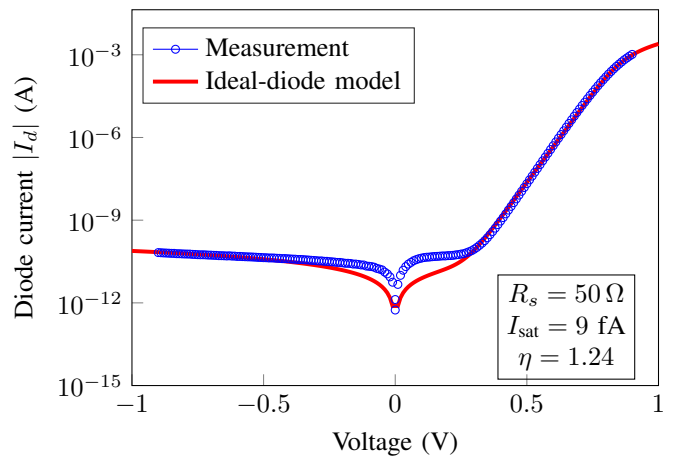}
\caption{Diode current-voltage characteristic. Comparison of the diode model and on-wafer dc-measurements of a fabricated $\times6$-harmonic Schottky diode integrated mixer circuit with a Schottky contact area of 0.11~$\mu$m$^2$. Extracted diode parameters are as shown. The measurements were performed with Kelvin probes at room temperature (dark condition). \textcolor{black}{The applied dc-voltage was swept from $-0.9$ to +0.9~V as the maximum current was limited to 1~mA.}}
\label{fig:IV}
\end{figure}

Before the release of the integrated mixer circuit from the supporting substrate, on-wafer, four-point dc measurements were performed using an Agilent B1500A semiconductor device parameter analyzer. The diode parameters such as the dc series resistance, ideality factor, and saturation current ($R_\text{s} = 50~\Omega$, $\eta = 1.24$, $I_{\text{sat}} = 9~$fA) were determined by fitting the measurement data to the diode model as shown in Fig.~\ref{fig:IV}. To model the substrate leakage effect, a shunt resistor of about 13~G$\Omega$ was added in parallel to the diode model. The forward breakdown voltage of the diode is around 1.1~V. 

\begin{figure*}[!t]
\centering
\includegraphics[width=\linewidth]{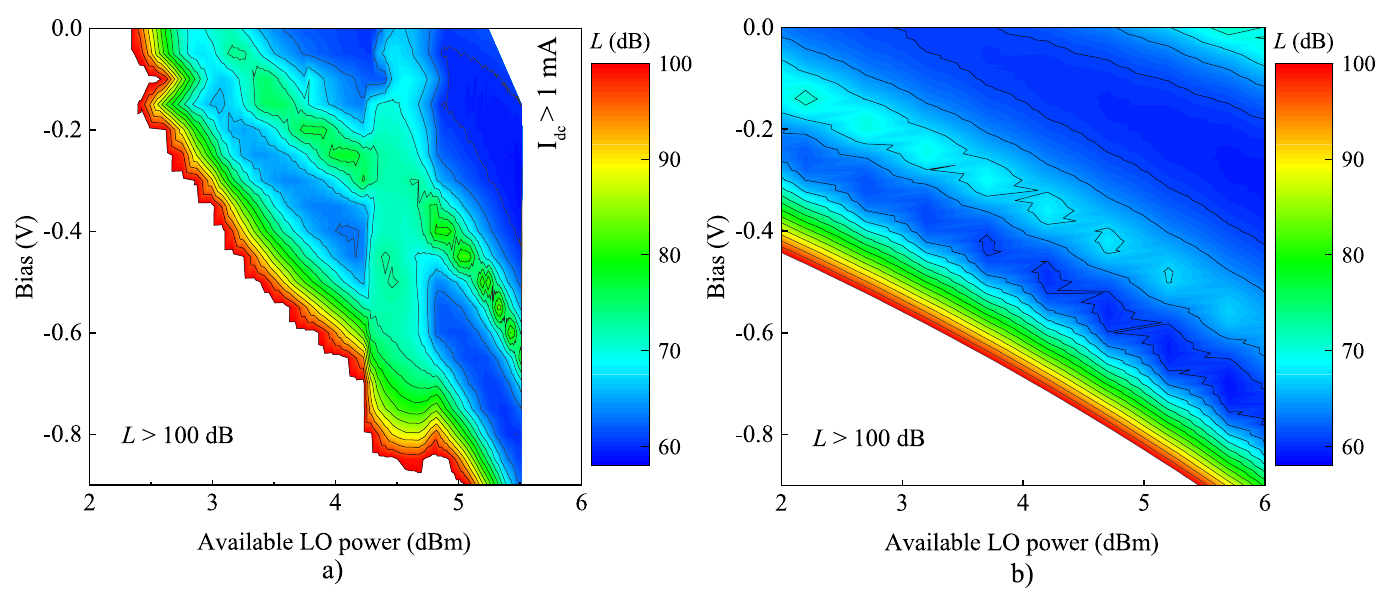}
\caption{Conversion loss as a function of LO power and bias. (a) Measurement: Conversion loss measured at an IF of 200 MHz, an RF of 3.443 THz, and for an LO signal at 573.3 GHz. The measured RF power at the mixer interface was about \SI{-3}{dBm} and the available LO power was about 5.6 dBm at 573 GHz. At low LO powers, the measurement is affected by the noise floor of the system, while at higher powers it is limited by the maximum available LO power. (b) Simulation: Harmonic-balance simulation with an ideal-diode model assigned with parameters extracted from the dc measurements. The contour step is 3~dB.}
\label{fig:cont}
\end{figure*}

\subsection{RF characterization}
The characterization of the 3.5-THz, $\times$6-harmonic mixer was performed using the setup shown in Fig.~\ref{setup1}. \textcolor{black}{The mixer conversion loss was then determined by taking the difference between the input RF power and IF output power. Losses in the cable as well as connectors and the gain of the low noise amplifier's (LNA) were considered.} 

\textcolor{black}{The absolute power measurement was carried out using a Thomas Keating power meter at the mixer interface thereby eliminating the need to de-embed the atmospheric attenuation at 3.5~THz. The THz beam from the QCL was amplitude modulated using a mechanical chopper with a frequency of 15 Hz. A TPX lens was used to focus the beam and the power meter head was aligned such that the incoming signal is incident at the Brewster angle. The frequency was determined with high precision ($<$ 1 MHz) by measuring the absorption of the QCL radiation guided through an absorption cell filled with methanol and comparing this spectrum with a reference spectrum of methanol.} The available input RF power at 3.443~THz was \SI{-3}{dBm}. The IF output power was read out from the spectrum analyzer. The input synthesizer frequency was tuned to set the LO frequency as 573.7~GHz, which resulted in a 200-MHz IF signal as shown in Fig.~\ref{fig:beat}. An SNR of 40~dB was achieved at an IF of 200~MHz as shown in Fig.~\ref{fig:beat}. The beat signal was recorded with a resolution bandwidth (RBW) of about 510~kHz and a video bandwidth of 10~kHz.

\begin{figure}[!t]
    \centering
\definecolor{mycolor1}{rgb}{0,0,1}%
\definecolor{mycolor2}{rgb}{0.85000,0.32500,0.09800}%
\begin{tikzpicture}

\begin{axis}[%
width=7cm,
height=5cm,
at={(0.758in,0.481in)},
scale only axis,
unbounded coords=jump,
xmin=-1,
xmax=0,
xlabel={Bias (V)},
ymin=40,
ymax=100,
ytick = {40,60,80,100},
xtick pos = left, 
ytick pos = left,
xtick align = inside,
ytick align = inside,
ylabel={Conversion loss (dB)},
axis background/.style={fill=white},
legend style={at={(0.55,0.775)}, anchor=south west, legend cell align=left, align=left, draw=white!15!black}
]
\addplot [color=mycolor1, line width=1.5pt, mark=o, mark size=2pt, mark options={solid, mycolor1} ]
  table[row sep=crcr]{%
-0	61.4392\\
-0.05	60.4758\\
-0.1	59.7035\\
-0.15	59.5323\\
-0.2	59.3059\\
-0.25	59.5469\\
-0.3	60.0332\\
-0.35	61.8005\\
-0.4	63.7073\\
-0.45	67.7355\\
-0.5	80.9433\\
-0.55	71.4373\\
-0.6	65.2506\\
-0.65	61.4451\\
-0.7	60.5677\\
-0.75	61.3686\\
-0.8	65.8493\\
-0.85	73.4191\\
-0.9	83.2719\\
};
\addlegendentry{Measurement}

\addplot [color=red, line width=2.0pt]
  table[row sep=crcr]{%
-0.00000000000000000E0	6.86746331454513914E1\\
-2.00000000000000000E-2	6.66378163011670743E1\\
-4.00000000000000000E-2	6.49677905870124128E1\\
-6.00000000000000000E-2	6.36315500239928689E1\\
-8.00000000000000000E-2	6.25653518720513180E1\\
-1.00000000000000000E-1	6.17146990912179394E1\\
-1.20000000000000018E-1	6.10385328539317040E1\\
-1.40000000000000013E-1	6.05072383887770471E1\\
-1.60000000000000009E-1	6.01001835507765847E1\\
-1.79999999999999982E-1	5.98036853464172680E1\\
-1.99999999999999978E-1	5.96093800157290410E1\\
-2.19999999999999973E-1	5.95128386929021147E1\\
-2.39999999999999947E-1	5.95122855549259366E1\\
-2.59999999999999964E-1	5.96073238208394685E1\\
-2.79999999999999982E-1	5.97976507624152376E1\\
-3.00000000000000000E-1	6.00818586055755954E1\\
-3.20000000000000018E-1	6.04565628050321813E1\\
-3.40000000000000036E-1	6.09162197783710990E1\\
-3.60000000000000053E-1	6.14539831676124493E1\\
-3.80000000000000071E-1	6.20636814688484595E1\\
-4.00000000000000089E-1	6.27424113774661407E1\\
-4.20000000000000107E-1	6.34922999540996624E1\\
-4.40000000000000124E-1	6.43183723536759544E1\\
-4.60000000000000142E-1	6.52158366261354683E1\\
-4.80000000000000160E-1	6.61321911996667744E1\\
-5.00000000000000089E-1	6.68854402751465038E1\\
-5.20000000000000107E-1	6.70936261536041734E1\\
-5.40000000000000124E-1	6.63963353090943276E1\\
-5.60000000000000142E-1	6.49290964762524858E1\\
-5.80000000000000160E-1	6.31949041651348153E1\\
-6.00000000000000178E-1	6.15913050777565818E1\\
-6.20000000000000195E-1	6.03177513052075653E1\\
-6.40000000000000213E-1	5.94740844777699262E1\\
-6.60000000000000231E-1	5.91390364511266142E1\\
-6.80000000000000249E-1	5.94086509563029885E1\\
-7.00000000000000266E-1	6.04014147049817129E1\\
-7.20000000000000284E-1	6.22263036693186944E1\\
-7.40000000000000302E-1	6.49221571956674381E1\\
-7.60000000000000320E-1	6.84137527842768201E1\\
-7.80000000000000338E-1	7.25325616938552020E1\\
-8.00000000000000355E-1	7.70848837518155428E1\\
-8.20000000000000462E-1	8.19080627504761516E1\\
-8.40000000000000391E-1	8.68887599770290642E1\\
-8.60000000000000497E-1	9.19563633492472832E1\\
-8.80000000000000426E-1	9.70696991237336704E1\\
-9.00000000000000533E-1	1.02205730793272864E2\\
-9.20000000000000462E-1	1.07352002574944594E2\\
-9.40000000000000568E-1	1.12502111465038057E2\\
-9.60000000000000497E-1	1.17653151018122060E2\\
-9.80000000000000604E-1	1.22804318308965921E2\\
-1.00000000000000000E0	1.27956142428358355E2\\
};
\addlegendentry{Model}

\end{axis}
\end{tikzpicture}%

%0.95	84.1985634843492\\
%1	96.86969183925\\
    \caption{Model fit. A vertical cross section from the measurement and simulation contours presented in Fig.~\ref{fig:cont}. Conversion loss plotted versus dc bias for an LO pump power of about 5.2~dBm. A conversion null feature arising due to destructive interference is observed at 0.5 V. }
    \label{ADSbias}
\end{figure}
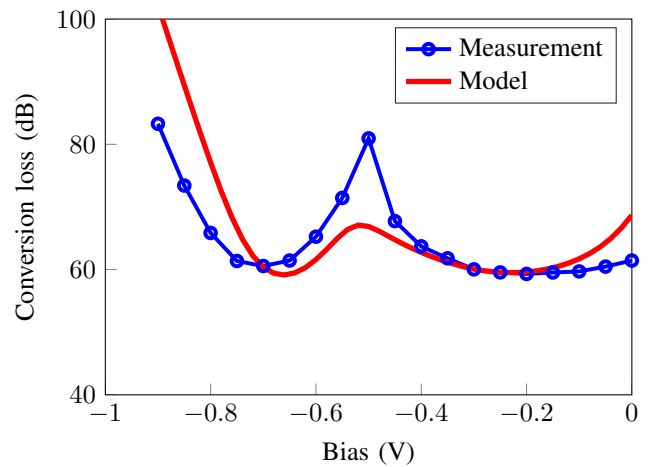

\begin{figure} [!b]
    \centering
    \input{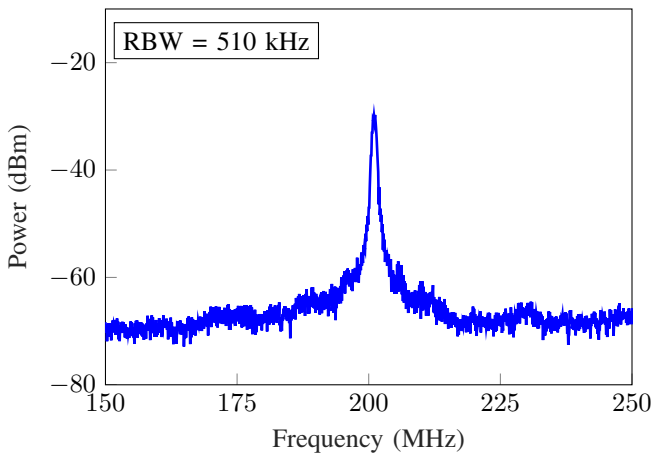}
    \caption{Beat signal. The down-converted IF signal generated at 200 MHz by the 3.5-THz, $\times$6-harmonic, single-ended Schottky diode mixer. }
    \label{fig:beat}
\end{figure}

\begin{table*}[!t]
\caption{\textcolor{black}{State-of-the-art mixers available for phase locking of QCLs}}
\label{table3}
\begin{tabularx}{\textwidth}{@{}l*{10}{C}c@{}}
\textbf{Frequency} & \textbf{Technology} & \textbf{Harmonic number}  & \textbf{Conversion loss}   & \textbf{SNR}  & \textbf{Resolution bandwidth} & \textbf{Reference} \\ 
 \toprule
 \SI{2.32}{THz} & Balanced Schottky-diode mixer  & $\times 21$ & $\SI{\sim 110}{dB}$  & \SI{25}{dB}  & \SI{1}{kHz}& \cite{Danylov2015} \\
 \midrule
 \SI{2.69}{THz} & Single-ended Schottky diode mixer & $\times 4$ & \SI{30}{dB} &  &  & \cite{bulcha2016}\\
 \midrule
\textcolor{black}{\SI{2.7}{THz}} & \textcolor{black}{Superlattice diode mixer} & \textcolor{black}{$\times 15$} & & \textcolor{black}{\SI{15}{dB}} & \textcolor{black}{\SI{3}{kHz}} & \textcolor{black}{\cite{Khosropanah2009}} \\
 \midrule
 \SI{3.4}{THz} & Superlattice diode mixer & $\times 18$ & $\SI{\sim 80}{dB}$  & \SI{30}{dB}  & \SI{100}{kHz} & \cite{hayton2013} \\ 
 \midrule
 $\mathbf{3.5 THz}$ & \textbf{Single-ended Schottky diode mixer} & $\mathbf{\times 6}$ &  $\mathbf{59~dB}$ & $\mathbf{40~dB}$  & $\mathbf{510~kHz}$ & \textbf{This work} \\
 \midrule
 \SI{4.7}{THz} & Cooled (10 K) superlattice diode mixer & $\times 24$ &  & \SI{20}{dB} &   \SI{30}{kHz} & \cite{khudchenko} \\
\bottomrule
\end{tabularx}
\end{table*} 

In Fig.~\ref{fig:cont}(a), the measured conversion loss of the mixer versus available LO pump power and dc bias is illustrated as a contour plot. The 1-mA current safety limit of the diode defines the upper limit for the dc bias and LO power. The noise in the IF power measurement sets an upper limit of the measurable conversion loss of about 100~dB. A conversion loss roll-off followed by a valley and null is observed. Strong features referred to as conversion nulls appearing at specific bias points for a constant LO power is caused by destructive interference arising from mixing products \cite{Feinaugle2002}. The conversion loss saturates at approximately 59~dB for an LO power higher than 5~dBm. 

To understand the mixer performance, a large-signal simulation was performed, and the results are shown in Fig.~\ref{fig:cont}(b). For this simulation, the ideal-diode model described in Section~\ref{sectionIIA} was modified to obtain a good agreement with the measurement results. In addition, a field solution was generated inside the lossy ohmic contact mesa for an accurate simulation of induced currents, which resulted in an additional loss of 5~dB \cite{Tang2011}. Finally, by assuming a loss of 8 dB on the RF side and taking into account a conductor loss of approximately 1~dB in the 11-mm-long LO WM-380 access waveguide in the mixer, we were able to reproduce the measured results of conversion loss versus bias for a constant LO power (5.2 dBm) as shown in Fig.~\ref{ADSbias}. Still, we were not able to fully reproduce the measured conversion loss versus LO power due to the strong interaction created by the large standing waves between the mixer and the last LO multiplier stage.

%The diode parameters extracted from the dc measurements ($R_\text{s} = 50~\Omega$, $\eta = 1.24$, $I_{\text{sat}} = 9~$fA) and a 13-G$\Omega$ shunt resistor were added to the diode model. 

\textcolor{black}{The 8-dB loss on the RF side has several contributors: optical coupling loss between the QCL beam and the RF diagonal feed horn \cite{goldsmith}, and conductor loss in the waveguide}. However, we believe that a significant part of the discrepancy between model and measurement could be due to possible air gaps in the split plane of the mixer block and substrate loss in the GaAs membrane. At high LO power levels, self-heating effects will contribute to additional loss due to a rather high thermal resistance between the Schottky contact and the waveguide housing. In addition, block fabrication tolerances could result in a degradation as shown in the sensitivity analysis in Section~\ref{sectionIIC}. For instance, the relative permittivity of the GaAs membrane is a critical parameter, which is frequency dependent and expected to be slightly higher at 3.5 THz compared to the low frequency asymptote of $\epsilon_s = 12.9$ \cite{Grischkowsky}. 

The results presented in this work are summarized along with other state-of-the-art technologies available for QCL frequency stabilization in Table~\ref{table3}. Although the presented 3.5-THz, $\times$6-harmonic mixer exhibits a higher conversion loss than in the model, the SNR of the IF signal is sufficient \textcolor{black}{for phase locking both 3.5-THz and 4.7 THz QCLs \cite{HeikoIRMMW}.}

\section{Conclusion}

In this paper, we have demonstrated a 3.5-THz, $\times$6-harmonic Schottky diode mixer operating at room temperature for frequency stabilization of a 3.5-THz QCL. We achieved a mixer conversion loss of about 59 dB for an IF signal of 200 MHz providing an SNR better than 40~dB which is more than sufficient to stabilize the QCL to a reference oscillator. The design can be further improved by making sure that the mixer performance is immune to the circuit and block fabrication tolerances especially on the RF side \cite{byron2021} as well as reducing the RF waveguide loss. We also observed that it is essential to optimize the diode geometry \cite{bobby} in order to reduce high-frequency losses, which contribute to the spreading series resistance. For the future, thermal management and the influence of self-heating effects on mixer performance should be addressed. The results show promising prospect for the availability of compact, uncooled, and efficient THz harmonic mixers, which opens up a plethora of opportunities for building air/space-borne heterodyne receivers with a high spectral resolution.

\section*{Acknowledgment}
The authors wish to acknowledge Dr. E. Saenz, ESTEC, European Space Agency (ESA), Noordwijk, The Netherlands, for excellent supervision of the project and for fruitful discussions. The authors would also like to thank M. Myremark for precision machining of mixer split-blocks, Dr. J. Vukusic for his assistance with gold sputtering on the waveguide housing, Dr. K. Biermann for the growth of the QCLs, as well as W. Anders, A. Riedel, and Dr. A. Tahraoui for sample preparation. 

\bibliographystyle{IEEEtran}
\bibliography{IEEEfull,bibl}

\begin{thebibliography}{10}
\providecommand{\url}[1]{#1}
\csname url@samestyle\endcsname
\providecommand{\newblock}{\relax}
\providecommand{\bibinfo}[2]{#2}
\providecommand{\BIBentrySTDinterwordspacing}{\spaceskip=0pt\relax}
\providecommand{\BIBentryALTinterwordstretchfactor}{4}
\providecommand{\BIBentryALTinterwordspacing}{\spaceskip=\fontdimen2\font plus
\BIBentryALTinterwordstretchfactor\fontdimen3\font minus
  \fontdimen4\font\relax}
\providecommand{\BIBforeignlanguage}[2]{{%
\expandafter\ifx\csname l@#1\endcsname\relax
\typeout{** WARNING: IEEEtran.bst: No hyphenation pattern has been}%
\typeout{** loaded for the language `#1'. Using the pattern for}%
\typeout{** the default language instead.}%
\else
\language=\csname l@#1\endcsname
\fi
#2}}
\providecommand{\BIBdecl}{\relax}
\BIBdecl

\bibitem{Waters2006}
J.~W. Waters, L.~Froidevaux, R.~S. Harwood, R.~F. Jarnot, H.~M. Pickett, W.~G.
  Read, P.~H. Siegel, R.~E. Cofield, M.~J. Filipiak, D.~A. Flower, J.~R.
  Holden, G.~K. Lau, N.~J. Livesey, G.~L. Manney, H.~C. Pumphrey, M.~L. Santee,
  D.~L. Wu, D.~T. Cuddy, R.~R. Lay, M.~S. Loo, V.~S. Perun, M.~J. Schwartz,
  P.~C. Stek, R.~P. Thurstans, M.~A. Boyles, K.~M. Chandra, M.~C. Chavez, G.-S.
  Chen, B.~V. Chudasama, R.~Dodge, R.~A. Fuller, M.~A. Girard, J.~H. Jiang,
  Y.~Jiang, B.~W. Knosp, R.~C. LaBelle, J.~C. Lam, K.~A. Lee, D.~Miller, J.~E.
  Oswald, N.~C. Patel, D.~M. Pukala, O.~Quintero, D.~M. Scaff, W.~Van~Snyder,
  M.~C. Tope, P.~A. Wagner, and M.~J. Walch, ``{The Earth Observing System
  Microwave Limb Sounder ({EOS MLS}) on the Aura Satellite},'' \emph{IEEE
  Trans. Geosci. Remote Sens.}, vol.~44, no.~5, pp. 1075--1092, May 2006,
  {DOI}:~\href{https://doi.org/10.1109/TGRS.2006.873771}{10.1109/TGRS.2006.873771}.

\bibitem{Richter2021}
H.~Richter, C.~Buchbender, R.~G\"{u}sten, R.~Higgins, B.~Klein, J.~Stutzki,
  H.~Wiesemeyer, and H.-W. H\"{u}bers, ``Direct measurements of atomic oxygen
  in the mesosphere and lower thermosphere using terahertz heterodyne
  spectroscopy,'' \emph{Commun. Earth Environment}, vol.~2, no.~1, Jan. 2021,
  {Art.} no. 19,
  {DOI}:~\href{https://doi.org/10.1038/s43247-020-00084-5}{10.1038/s43247-020-00084-5}.

\bibitem{Siegel2007}
P.~H. {Siegel}, ``{THz Instruments for Space},'' \emph{IEEE Trans. Antennas
  Propag.}, vol.~55, no.~11, pp. 2957--2965, Nov. 2007,
  {DOI}:~\href{https://doi.org/10.1109/TAP.2007.908557}{10.1109/TAP.2007.908557}.

\bibitem{Farrah2019}
D.~Farrah, K.~E. Smith, D.~Ardila, C.~M. Bradford, M.~Dipirro, C.~Ferkinhoff,
  J.~Glenn, P.~Goldsmith, D.~Leisawitz, T.~Nikola, N.~Rangwala, S.~A. Rinehart,
  J.~Staguhn, M.~Zemcov, J.~Zmuidzinas, J.~Bartlett, S.~Carey, W.~J. Fischer,
  J.~Kamenetzky, J.~Kartaltepe, M.~Lacy, D.~C. Lis, L.~Locke,
  E.~Lopez-Rodriguez, M.~MacGregor, E.~Mills, S.~H. Moseley, E.~J. Murphy,
  A.~Rhodes, M.~Richter, D.~Rigopoulou, D.~Sanders, R.~Sankrit, G.~Savini,
  J.-D. Smith, and S.~Stierwalt, ``Review: far-infrared instrumentation and
  technological development for the next decade,'' \emph{J. Astron. Telesc.
  Instrumen. Syst.}, vol.~5, Apr. 2019, {Art.} no. 020901,
  {DOI}:~\href{https://doi.org/10.1117/1.jatis.5.2.020901}{10.1117/1.jatis.5.2.020901}.

\bibitem{Koehler2002}
R.~K\"{o}hler, A.~Tredicucci, F.~Beltram, H.~E. Beere, E.~H. Linfield, A.~G.
  Davies, D.~A. Ritchie, R.~C. Iotti, and F.~Rossi, ``Terahertz
  semiconductor-heterostructure laser,'' \emph{Nature}, vol. 417, pp. 156--159,
  May 2002, {DOI}:~\href{https://doi.org/10.1038/417156a}{10.1038/417156a}.

\bibitem{Richter2015}
H.~Richter, M.~Wienold, L.~Schrottke, K.~Biermann, H.~T. Grahn, and H.-W.
  Hübers, ``4.7-{THz} local oscillator for the {GREAT} heterodyne spectrometer
  on {SOFIA},'' \emph{IEEE Trans. THz Sci. Technol.}, vol.~5, no.~4, pp.
  539--545, Jul. 2015,
  {DOI}:~\href{https://doi.org/10.1109/TTHZ.2015.2442155}{10.1109/TTHZ.2015.2442155}.

\bibitem{Kloosterman2013}
\textcolor{black}{J. L. Kloosterman, D. J. Hayton, Y. Ren, T. Y. Kao, J. N.
  Hovenier, J. R. Gao, T. M. Klapwijk, Q. Hu, C. K. Walker, and J. L. Reno},
  ``\textcolor{black}{Hot electron bolometer heterodyne receiver with a
  4.7-{THz} quantum cascade laser as a local oscillator},''
  \emph{\textcolor{black}{Appl. Phys. Lett.}}, vol. \textcolor{black}{102}, no.
  \textcolor{black}{1}, Jan. \textcolor{black}{2013}, \textcolor{black}{{Art.}
  no. 011123,
  {DOI}:~\href{https://doi.org/10.1063/1.4774085}{10.1063/1.4774085}}.

\bibitem{Betz}
A.~L. Betz, R.~T. Boreiko, B.~S. Williams, S.~Kumar, Q.~Hu, and J.~L. Reno,
  ``Frequency and phase-lock control of a 3 {THz} quantum cascade laser,''
  \emph{Opt. Lett.}, vol.~30, no.~14, pp. 1837--1839, Jul. 2005,
  {DOI}:~\href{https://doi.org/10.1364/OL.30.001837}{10.1364/OL.30.001837}.

\bibitem{Rabanus2009}
D.~Rabanus, U.~U. Graf, M.~Philipp, O.~Ricken, J.~Stutzki, B.~Vowinkel, M.~C.
  Wiedner, C.~Walther, M.~Fischer, and J.~Faist, ``Phase locking of a 1.5
  terahertz quantum cascade laser and use as a local oscillator in a heterodyne
  {HEB} receiver,'' \emph{Opt. Express}, vol.~17, no.~3, pp. 1159--1168, Feb.
  2009,
  {DOI}:~\href{https://doi.org/10.1364/oe.17.001159}{10.1364/oe.17.001159}.

\bibitem{Khosropanah2009}
\textcolor{black}{P. Khosropanah, A. Baryshev, W. Zhang, W. Jellema, J. N.
  Hovenier, J. R. Gao, T. M. Klapwijk, D. G. Paveliev, B. S. Williams, S.
  Kumar, Q. Hu, J. L. Reno, B. Klein, and J. L. Hesler},
  ``\textcolor{black}{Phase locking of a 2.7 {THz} quantum cascade laser to a
  microwave reference},'' \emph{\textcolor{black}{Opt. Lett.}}, vol.~34,
  no.~19, pp. \textcolor{black}{2958--2060}, Oct. \textcolor{black}{2009},
  \textcolor{black}{{DOI}:~\href{https://doi.org/10.1364/ol.34.002958}{10.1364/ol.34.002958}}.

\bibitem{hayton2013}
D.~J. Hayton, A.~Khudchencko, D.~G. Pavelyev, J.~N. Hovenier, A.~Baryshev,
  J.~R. Gao, T.~Y. Kao, Q.~Hu, J.~L. Reno, and V.~Vaks, ``Phase locking of a
  3.4 {THz} third-order distributed feedback quantum cascade laser using a
  room-temperature superlattice harmonic mixer,'' \emph{Appl. Phys. Lett.},
  vol. 103, Aug. 2013, {Art.} no. 051115,
  {DOI}:~\href{https://doi.org/10.1063/1.4817319}{10.1063/1.4817319}.

\bibitem{khudchenko}
A.~V. Khudchenko, D.~J. Hayton, D.~G. Pavelyev, A.~M. Baryshev, J.~R. Gao,
  T.-Y. Kao, Q.~Hu, J.~L. Reno, and V.~L. Vaks, ``{Phase locking a 4.7 {THz}
  quantum cascade laser using a super-lattice diode as harmonic mixer},'' in
  \emph{2014 39th Int. Conf. on Infrared, Millimeter, and Terahertz waves
  (IRMMW-THz)}, 2014, pp. 1--2,
  {DOI}:~\href{https://doi.org/10.1109/IRMMW-THz.2014.6956455}{10.1109/IRMMW-THz.2014.6956455}.

\bibitem{Danylov2015}
A.~Danylov, N.~Erickson, A.~Light, and J.~Waldman, ``Phase locking of 2.324 and
  2.959 terahertz quantum cascade lasers using a {Schottky} diode harmonic
  mixer,'' \emph{Opt. Lett.}, vol.~40, no.~21, pp. 5090--5092, Nov. 2015,
  {DOI}:~\href{https://doi.org/10.1364/ol.40.005090}{10.1364/ol.40.005090}.

\bibitem{bulcha2016}
B.~T. {Bulcha}, J.~L. {Hesler}, V.~{Drakinskiy}, J.~{Stake}, A.~{Valavanis},
  P.~{Dean}, L.~H. {Li}, and N.~S. {Barker}, ``Design and characterization of
  1.8–3.2 {THz} {Schottky}-based harmonic mixers,'' \emph{IEEE Trans. THz
  Sci. Technol.}, vol.~6, no.~5, pp. 737--746, Sep. 2016,
  {DOI}:~\href{https://doi.org/10.1109/TTHZ.2016.2576686}{10.1109/TTHZ.2016.2576686}.

\bibitem{imran}
I.~Mehdi, J.~V. Siles, C.~Lee, and E.~Schlecht, ``{THz} diode technology:
  Status, prospects, and applications,'' \emph{Proc. IEEE}, vol. 105, no.~6,
  pp. 990--1007, Jun. 2017,
  {DOI}:~\href{https://doi.org/10.1109/JPROC.2017.2650235}{10.1109/JPROC.2017.2650235}.

\bibitem{Waveguide}
``{IEEE Standard for Rectangular Metallic Waveguides and Their Interfaces for
  Frequencies of {110 GHz} and Above--Part 1: Frequency Bands and Waveguide
  Dimensions},'' \emph{{IEEE} Std 1785.1-2012}, pp. 1--22, Mar. 2013,
  {DOI}:~\href{https://doi.org/10.1109/IEEESTD.2013.6471987}{10.1109/IEEESTD.2013.6471987}.

\bibitem{crowe1992}
T.~W. {Crowe}, R.~J. {Mattauch}, H.~P. {Röser}, W.~L. {Bishop}, W.~C.~B.
  {Peatman}, and X.~{Liu}, ``{GaAs {Schottky} diodes for THz mixing
  applications},'' \emph{Proc. IEEE}, vol.~80, no.~11, pp. 1827--1841, Nov.
  1992, {DOI}:~\href{https://doi.org/10.1109/5.175258}{10.1109/5.175258}.

\bibitem{dickens}
L.~E. Dickens, ``Spreading resistance as a function of frequency,'' \emph{IEEE
  Trans. Microw. Theory Techn.}, vol.~15, no.~2, pp. 101--109, Feb. 1967,
  {DOI}:~\href{https://doi.org/10.1109/TMTT.1967.1126383}{10.1109/TMTT.1967.1126383}.

\bibitem{Tang2012}
A.~Y. {Tang}, E.~{Schlecht}, R.~{Lin}, G.~{Chattopadhyay}, C.~{Lee}, J.~{Gill},
  I.~{Mehdi}, and J.~{Stake}, ``Electro-thermal model for multi-anode
  {Schottky} diode multipliers,'' \emph{IEEE Trans. THz Sci. Technol.}, vol.~2,
  no.~3, pp. 290--298, May 2012,
  {DOI}:~\href{https://doi.org/10.1109/TTHZ.2012.2189913}{10.1109/TTHZ.2012.2189913}.

\bibitem{Diego2016}
\textcolor{black}{D. Moro-Melgar, A. Maestrini, J. Treuttel, L. Gatilova, T.
  González, B. Vasallo, and J. Mateos}, ``{\textcolor{black}{Monte Carlo Study
  of 2-D Capacitance Fringing Effects in GaAs Planar Schottky Diodes}},''
  \emph{\textcolor{black}{IEEE Trans. Electron Device Lett.}}, vol.
  \textcolor{black}{63}, no. \textcolor{black}{10}, pp.
  \textcolor{black}{3900--3907}, Oct. \textcolor{black}{2016},
  \textcolor{black}{{DOI}:~\href{https://doi.org/10.1109/TED.2016.2601341}{10.1109/TED.2016.2601341}},.

\bibitem{louhi1994}
J.~Louhi, ``{The capacitance of a small circular Schottky diode for
  submillimeter wavelengths},'' \emph{IEEE Microw. Guided Wave Lett.}, vol.~4,
  no.~4, pp. 107--108, Apr. 1994,
  {DOI}:~\href{https://doi.org/10.1109/75.282574}{10.1109/75.282574}.

\bibitem{Saleh}
A.~A.~M. Saleh, ``Theory of resistive mixers,'' Ph.D. dissertation, Dept.
  Elect. Eng., MIT, Cambridge, MA, USA, 1970.

\bibitem{Divya2019}
{D. Jayasankar, J. Stake, and P. Sobis}, ``Effect of idler terminations on the
  conversion loss for {THz} {Schottky} diode harmonic mixers,'' \emph{2019 44th
  Int. Conf. on Infrared, Millimeter, and Terahertz waves (IRMMW-THz)}, 2019,
  {Paris}, pp. 1--2,
  {DOI}:~\href{https://doi.org/10.1109/IRMMW-THz.2019.8873938}{10.1109/IRMMW-THz.2019.8873938}.

\bibitem{Laman2008}
N.~Laman and D.~Grischkowsky, ``Terahertz conductivity of thin metal films,''
  \emph{Appl. Phys. Lett.}, vol.~93, Aug. 2008, {Art.} no. 051105,
  {DOI}:~\href{https://doi.org/10.1063/1.2968308}{10.1063/1.2968308}.

\bibitem{Palik1997}
E.~D. Palik, ``Gallium arsenide ({GaAs}),'' in \emph{Handbook of Optical
  Constants of Solids}, E.~D. Palik, Ed.\hskip 1em plus 0.5em minus 0.4em\relax
  Boston, MA, USA: Academic Press, 1985, pp. 429--443,
  {DOI}:~\href{https://www.sciencedirect.com/science/article/pii/B9780080547213500204}{10.1016/B978-0-08-054721-3.50020-4}.

\bibitem{Grischkowsky}
D.~Grischkowsky, S.~Keiding, M.~van Exter, and {\relax Ch}.~Fattinger,
  ``Far-infrared time-domain spectroscopy with terahertz beams of dielectrics
  and semiconductors,'' \emph{J. Opt. Soc. Am. B}, vol.~7, no.~10, pp.
  2006--2015, Oct. 1990,
  {DOI}:~\href{https://doi.org/10.1364/JOSAB.7.002006}{10.1364/JOSAB.7.002006}.

\bibitem{horn}
J.~F. {Johansson} and N.~D. {Whyborn}, ``{The diagonal horn as a sub-millimeter
  wave antenna},'' \emph{IEEE Trans. Microw. Theory Techn.}, vol.~40, no.~5,
  pp. 795--800, May 1992,
  {DOI}:~\href{https://doi.org/10.1109/22.137380}{10.1109/22.137380}.

\bibitem{Smith2007}
E.~D. Smith, F.~Szidarovszky, W.~J. Karnavas, and A.~T. Bahill, ``Sensitivity
  analysis, a powerful system validation technique,'' \emph{Open Cybern. Syst.
  J.}, vol.~2, pp. 39--56, Jan. 2008,
  {DOI}:~\href{https://doi.org/10.2174/1874110x00802010039}{10.2174/1874110x00802010039}.

\bibitem{siegel1999}
P.~H. {Siegel}, R.~P. {Smith}, M.~C. {Graidis}, and S.~C. {Martin},
  ``{2.5-{THz} {GaAs} monolithic membrane-diode mixer},'' \emph{IEEE Trans.
  Microw. Theory Techn.}, vol.~47, no.~5, pp. 596--604, May 1999,
  {DOI}:~\href{https://doi.org/10.1109/22.763161}{10.1109/22.763161}.

\bibitem{Schrottke2016}
L.~Schrottke, X.~L\"{u}, G.~Rozas, K.~Biermann, and H.~T. Grahn, ``Terahertz
  {GaAs}/{AlAs} quantum-cascade lasers,'' \emph{Appl. Phys. Lett.}, vol. 108,
  Mar. 2016,
  {DOI}:~\href{https://doi.org/10.1063/1.4943657}{10.1063/1.4943657}.

\bibitem{Schrottke2020}
L.~Schrottke, X.~L\"{u}, B.~R\"{o}ben, K.~Biermann, T.~Hagelschuer, M.~Wienold,
  H.-W. H\"{u}bers, M.~Hannemann, J.~H. van Helden, J.~R\"{o}pcke, and H.~T.
  Grahn, ``High-performance {GaAs/AlAs} terahertz quantum-cascade lasers for
  spectroscopic applications,'' \emph{IEEE Trans. THz Sci. Technol.}, vol.~10,
  no.~2, pp. 133--140, Mar. 2020,
  {DOI}:~\href{https://doi.org/0.1109/TTHZ.2019.2957456}{0.1109/TTHZ.2019.2957456}.

\bibitem{DivyaIRMMW}
\textcolor{black}{{D. Jayasankar}, {V. Drakinskiy}, {P. Sobis}, and {J.
  Stake}}, ``\textcolor{black}{Development of Supra-{THz} {Schottky} Diode
  Harmonic Mixers},'' \emph{\textcolor{black}{presented at the 2021 46th Int.
  Conf. on Infrared, Millimeter, and Terahertz waves (IRMMW-THz)}}, Aug.
  \textcolor{black}{2021}, \textcolor{black}{{Chengdu}, pp. 1--2}.

\bibitem{Feinaugle2002}
R.~Feinäugle, H.-W. H\"{u}bers, H.~P. Röser, and J.~L. Hesler, ``{On the
  effect of {IF} power nulls in {Schottky} diode harmonic mixers},'' \emph{IEEE
  Trans. Microw. Theory Techn.}, vol.~50, no.~1, pp. 134--142, Jan. 2002,
  {DOI}:~\href{https://doi.org/10.1109/22.981257}{10.1109/22.981257}.

\bibitem{Tang2011}
A.~Y. {Tang} and J.~{Stake}, ``Impact of eddy currents and crowding effects on
  high-frequency losses in planar {Schottky} diodes,'' \emph{IEEE Trans.
  Electron Device Lett.}, vol.~58, no.~10, pp. 3260--3269, Oct. 2011,
  {DOI}:~\href{https://doi.org/10.1109/TED.2011.2160724}{10.1109/TED.2011.2160724}.

\bibitem{goldsmith}
P.~F. {Goldsmith}, ``{{Quasi-optical techniques}},'' \emph{Proc. IEEE},
  vol.~80, no.~11, pp. 1729--1747, Nov. 1992,
  {DOI}:~\href{https://doi.org/10.1109/5.175252}{10.1109/5.175252}.

\bibitem{HeikoIRMMW}
\textcolor{black}{H. Richter, N. Rothbart, M. Wienold, X. L\"{u}, L. Schrottke,
  H.T. Grahn, D. Jayasankar, V. Drakinskiy, P. Sobis, J. Stake, and H-W.
  H\"{u}bers}, ``\textcolor{black}{Phase Locking Of {3.5-THz} and {4.7-THz}
  Quantum-Cascade Lasers using a {Schottky} Diode Harmonic Mixer},''
  \emph{\textcolor{black}{presented at the 2021 46th Int. Conf. on Infrared,
  Millimeter, and Terahertz waves (IRMMW-THz)}}, Aug. \textcolor{black}{2021},
  \textcolor{black}{{Chengdu}, pp. 1--2}.

\bibitem{byron2021}
C.~Viegas, J.~Powell, H.~Liu, H.~Sanghera, L.~Donoghue, P.~G. Huggard, and
  B.~Alderman, ``On-chip integrated backshort for relaxation of machining
  accuracy requirements in frequency multipliers,'' \emph{IEEE Microw. Wireless
  Compon. Lett.}, vol.~31, no.~2, pp. 188--191, Feb. 2021,
  {DOI}:~\href{https://doi.org/10.1109/LMWC.2020.3033440}{10.1109/LMWC.2020.3033440}.

\bibitem{bobby}
N.~Alijabbari, M.~F. Bauwens, and R.~M. Weikle, ``{Design and Characterization
  of Integrated Submillimeter-Wave Quasi-Vertical Schottky Diodes},''
  \emph{IEEE Trans. THz Sci. Technol.}, vol.~5, no.~1, pp. 73--80, Jan. 2015,
  {DOI}:~\href{https://doi.org/10.1109/TTHZ.2014.2361434}{10.1109/TTHZ.2014.2361434}.

\end{thebibliography}

\begin{IEEEbiography}[{\includegraphics[width=1in,height=1.5in,clip,keepaspectratio]{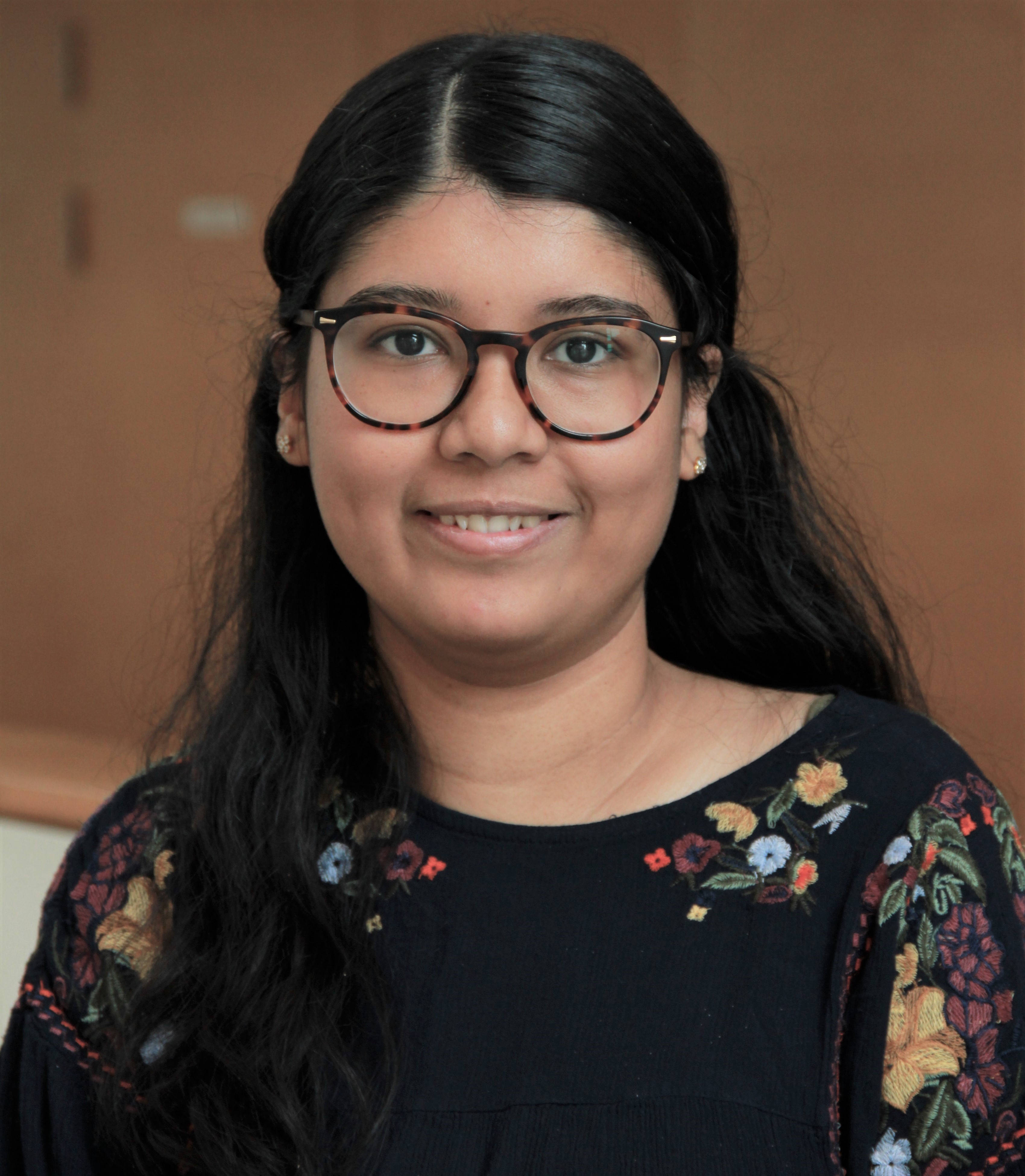}}]%
{Divya Jayasankar}
(S’19) was born in Madurai, India, in 1994. She received the bachelor degree in electronics and communication engineering from K.~S.~Rangasamy College of Technology, Tamilnadu, India, in 2015 and the M.Sc. degree in wireless, photonics and space engineering from the Chalmers University of Technology, Gothenburg, Sweden, in 2019.

From 2015 to 2017, she was with the Raman Research Institute, Bangalore, India, as a Research Assistant. In 2018, she joined the Terahertz and Millimeter Wave Laboratory, Department of Microtechnology and Nanoscience (MC2), Chalmers University of Technology, as a Project Assistant. She is currently employed as an institute doctoral student at the Research Institutes of Sweden and Chalmers University of Technology. Her research interests include the development of THz mixers for future space applications and THz metrology. 

D. Jayasankar received an internship award from the European Microwave Association (EuMA) in 2021 to join Prof. Emma Macpherson's group at the University of Warwick on THz in-vivo imaging.
\end{IEEEbiography}

\begin{IEEEbiography}[{\includegraphics[width=1in,height=1.5in,clip,keepaspectratio]{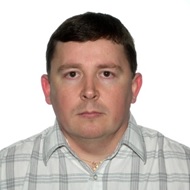}}]%
{Vladimir Drakinksiy} was born in Kurganinsk, Russia, in 1977. He received the Diploma degree in physics and informatics (with honors) from the Armavir State Pedagogical Institute, Armavir, Russia, in 2000 and the Postgraduate degree from Moscow State Pedagogical University, Moscow, Russia, in 2003. From 2000 to 2003, he was a Junior Research Assistant in the Physics Department, Moscow State Pedagogical University. Since 2003, he has been in the Department of Microtechnology and Nanoscience (MC2), Chalmers University of Technology, Gothenburg, Sweden. During 2003–2005, he was responsible for the mixer chip fabrication for the Herschel Space Observatory. Since 2008, he has been a Research Engineer with the MC2, Chalmers University of Technology. He is currently responsible for the terahertz Schottky diode process line at MC2, Chalmers University of Technology. His research interests include microfabrication and nanofabrication techniques, detectors for sub-millimeter and THz ranges, and superconducting thin films.
\end{IEEEbiography}

\begin{IEEEbiography}[{\includegraphics[width=1in,height=1.5in,clip,keepaspectratio]{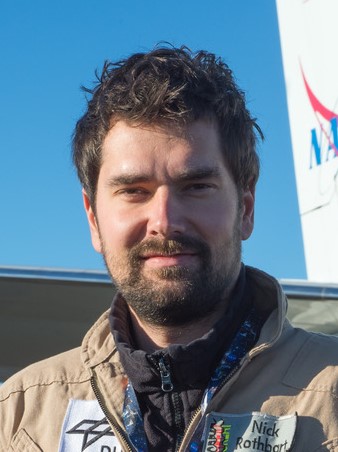}}]%
{Nick Rothbart} was born in Berlin, Germany, in 1985. He received the M.Sc. degree in physics from the Humboldt-Universität zu Berlin, Berlin, Germany, in 2011 and the Ph.D. (Dr. rer. nat.) degree in physics from the Technische Universität Berlin, Berlin, Germany, in 2015 for his involvement in THz imaging and spectroscopy that was accomplished at the German Aerospace Center (DLR), Berlin, Germany, and partially at the University of Massachusetts, Lowell, MA, USA, and supported by a scholarship and membership at Helmholtz Research School on Security Technologies. From 2014 to 2015, he was with the Federal Institute for Materials Research and Testing, Berlin, Germany. Since 2015, he has been involved in millimeter-wave/THz gas spectroscopy at DLR and the Humboldt-Universität zu Berlin.
\end{IEEEbiography}

\begin{IEEEbiography}[{\includegraphics[width=1in,height=1.5in,clip,keepaspectratio]{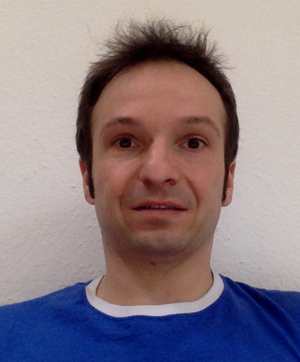}}]%
{Heiko Richter} received the Diplom in physics from the University of Karlsruhe, Karlsruhe, Germany, in 1999 and the Ph.D. degree in physics from the Technische Universität of Berlin, Berlin, Germany, in 2005. He is currently with the German Aerospace Center, Berlin, Germany, where he is involved in the field of THz and infrared sensors/optics. In 2007, he received the Lilienthal Award for the development of a THz security scanner.
\end{IEEEbiography}

\begin{IEEEbiography}[{\includegraphics[width=1in,height=1.5in,clip,keepaspectratio]{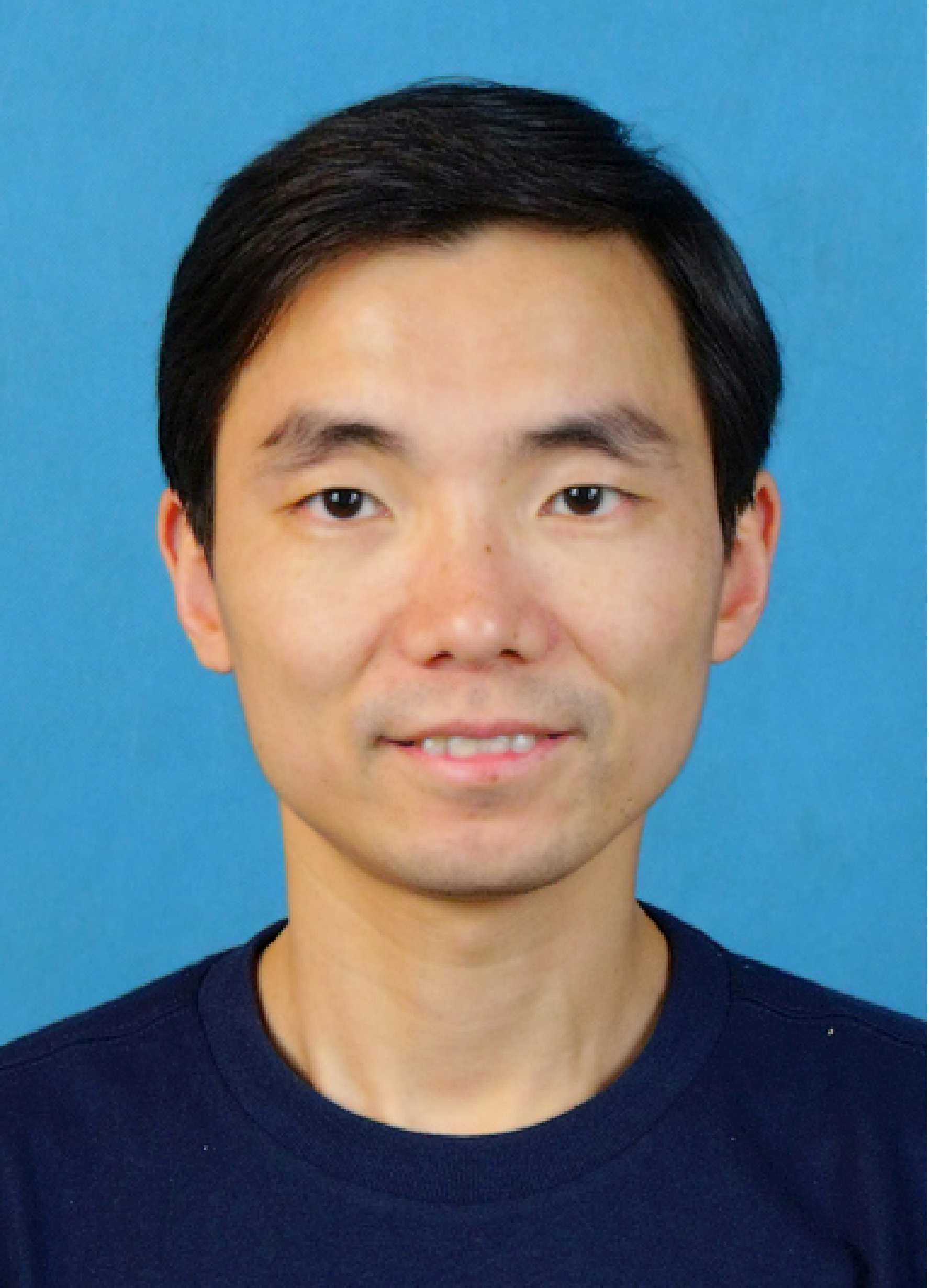}}]%
{Xiang L\"u}
received the Master degree in physics from Soochow University, Suzhou, China, in 2000 and the Ph.D. degree in physics from the Shanghai Institute for Technical Physics, Chinese Academy of Sciences, Shanghai, China, in 2003.
\newline
He is currently a Postdoctoral Research Associate at the Paul-Drude-Institut f\"ur Festk\"orperelektronik in Berlin, Germany, where he is involved in the field of THz quantum-cascade lasers.
\end{IEEEbiography}

\begin{IEEEbiography}[{\includegraphics[width=1in,height=1.5in,clip,keepaspectratio]{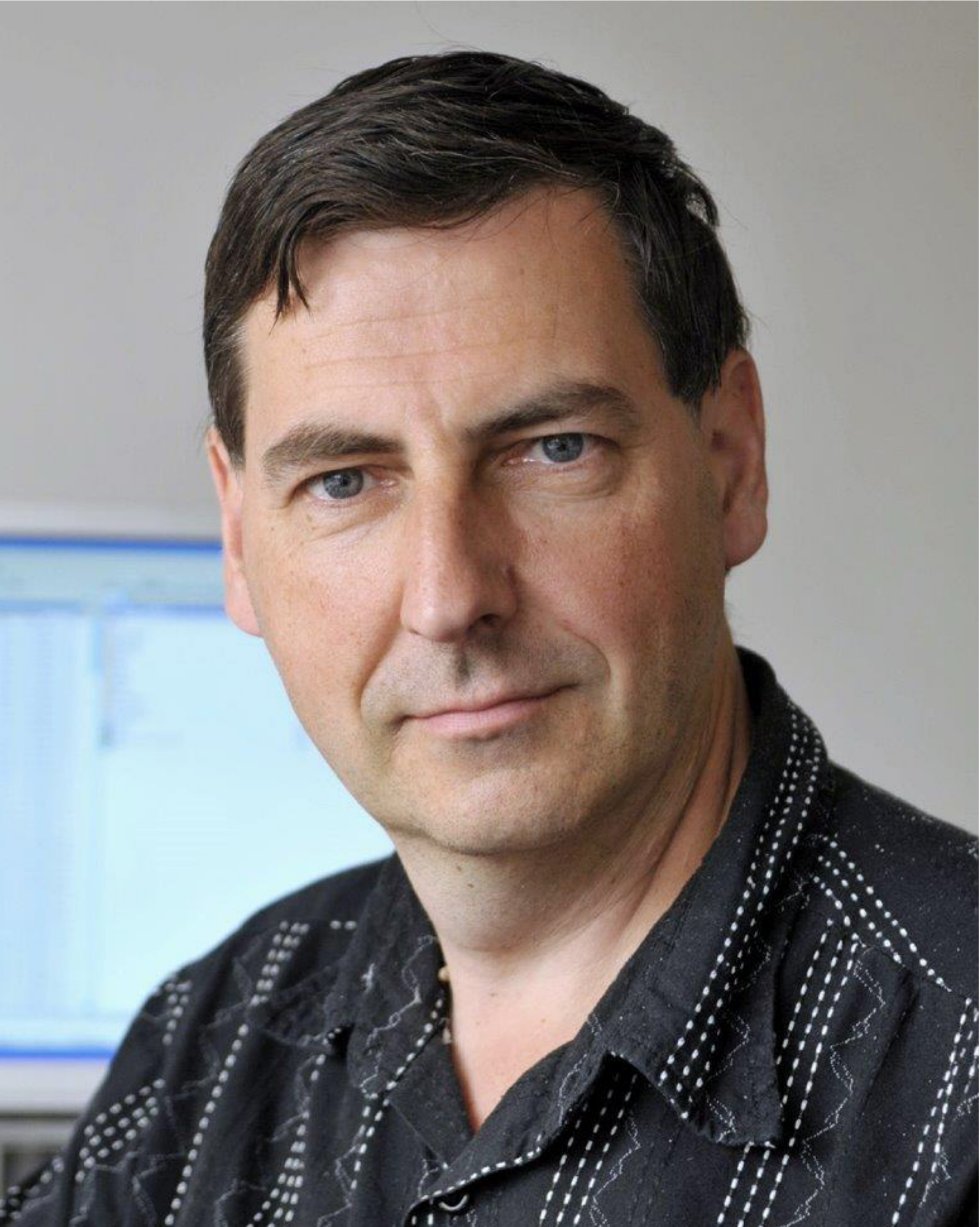}}]%
{Lutz Schrottke}
received the Diplom and doctoral degrees in {Experimental Physics} in 1983 and 1988, respectively, from the Humboldt-Universit\"at zu Berlin, Berlin, Germany.
\newline
From 1985 to 1991, he was with the Zentralinstitut f\"ur Elektronenphysik in Berlin, Germany, working on thin-film electroluminescent devices. In 1992, he joined the Paul-Drude-Institut f\"ur Festk\"orperelektronik
in Berlin, Germany, as a member of the Scientific {Staff}. He was a Visiting Scholar at the Department of Physics of the University of Michigan, Ann Arbor, MI, USA, from 1995 to 1996. His research interests include THz quantum-cascade lasers as well as optical and transport properties of semiconductor heterostructures.
\end{IEEEbiography}

\begin{IEEEbiography}[{\includegraphics[width=1in,height=1.5in,clip,keepaspectratio]{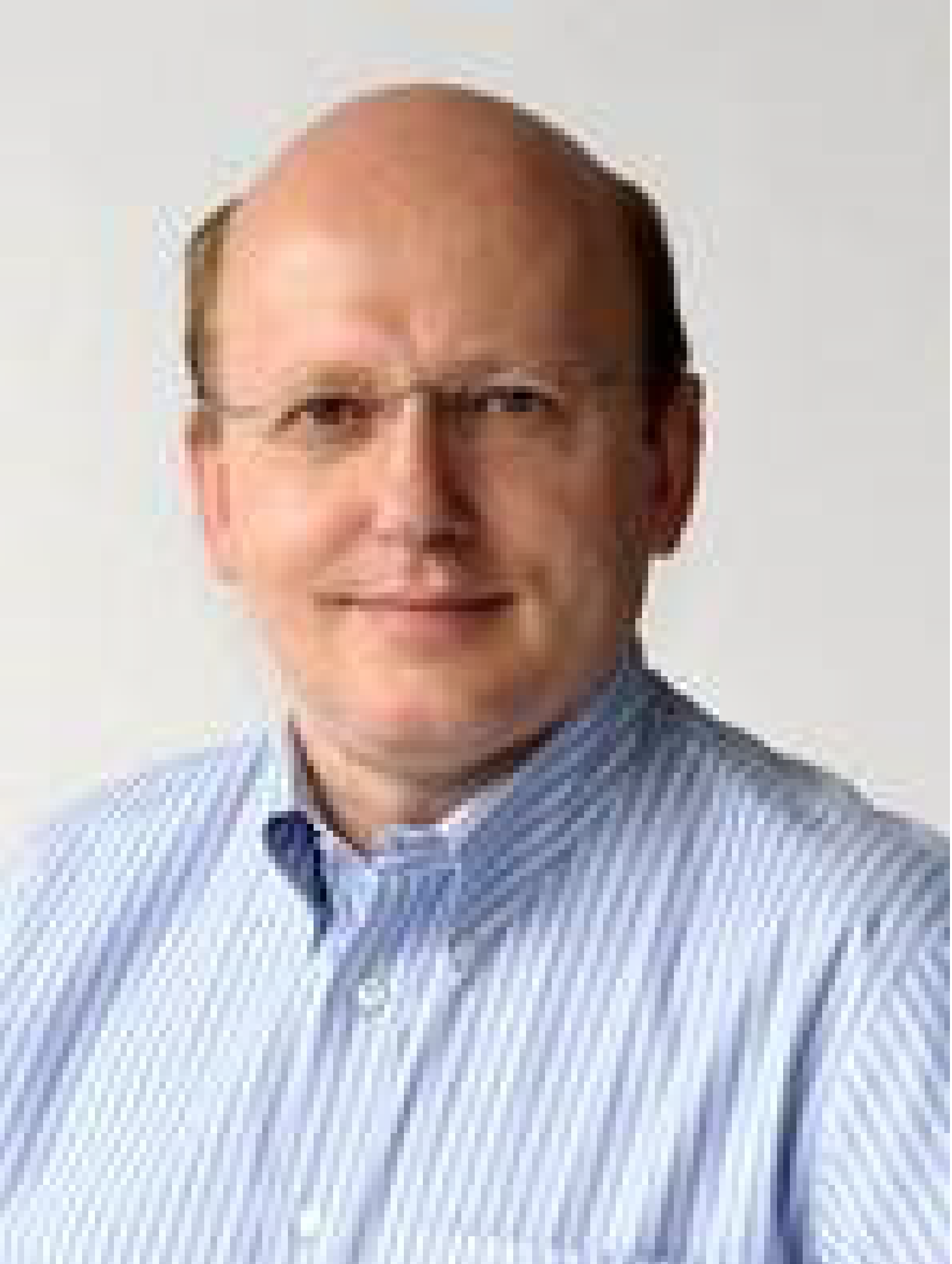}}]%
{Holger T. Grahn}
received the Diplom in physics from the Universit\"{a}t Kiel, Kiel, Germany, in 1983 and the Ph.D. degree in physics from Brown University, Providence, RI, USA, in 1987.
From 1988 to 1992, he was a Postdoctoral Research Assistant with the Max-Planck-Institut f\"{u}r Festk\"{o}rperforschung, Stuttgart, Germany, working on vertical transport in semiconductor superlattices. In 1992, he joined the Paul-Drude-Institut f\"{u}r Festk\"{o}rperelektronik, Berlin, Germany, as a Department Head, first for Analytics and later for Semiconductor Spectroscopy. In 2001, he was appointed an Adjunct Professor in physics at the Technische Universit\"{a}t Berlin, Berlin, Germany. His research interests include the optical and transport properties of semiconductor heterostructures and THz quantum-cascade lasers.
\end{IEEEbiography}

\begin{IEEEbiography}[{\includegraphics[width=1in,height=1.5in,clip,keepaspectratio]{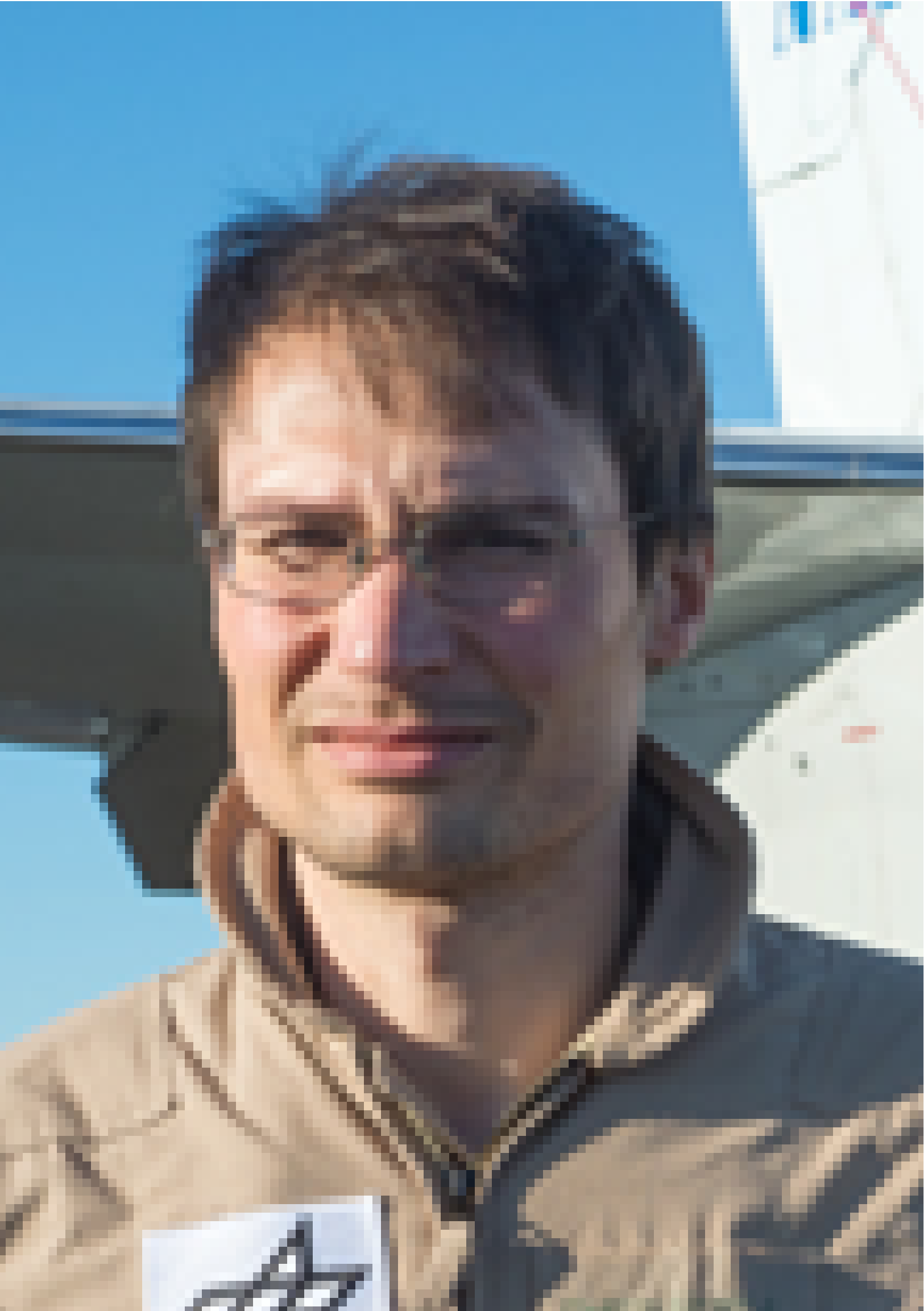}}]%
{Martin Wienold} received the Diplom and doctoral degree in physics from the Humboldt-Universität zu Berlin, Berlin, Germany, in 2007 and 2012, respectively, working on mid-infrared and THz quantum-cascade lasers. He is currently a Postdoctoral Research Associate with the German Aerospace Center, Berlin, Germany, working on the development of spectroscopic techniques based on THz quantum-cascade lasers.
\end{IEEEbiography}

\begin{IEEEbiography}[{\includegraphics[width=1in,height=1.5in,clip,keepaspectratio]{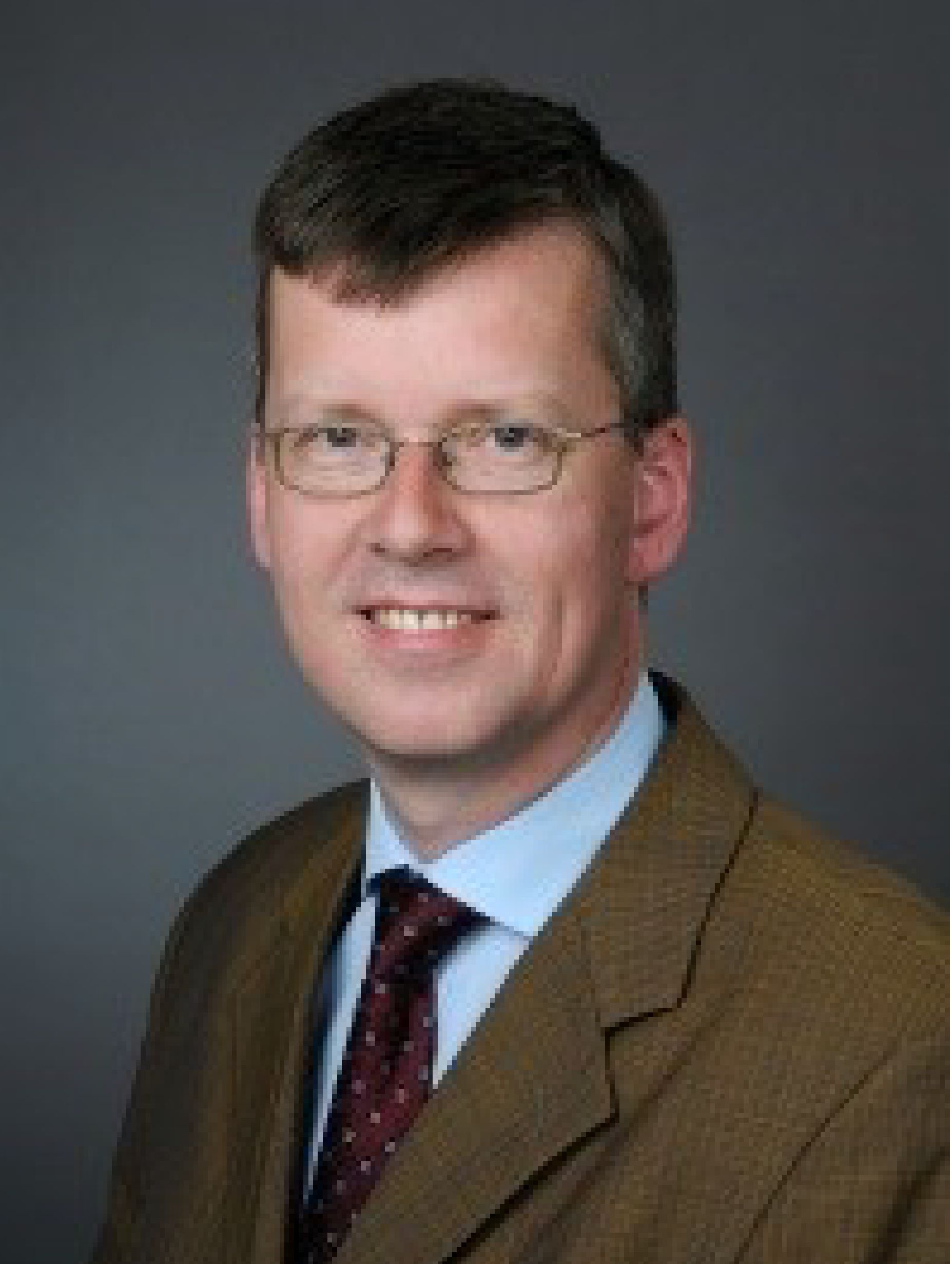}}]%
{Heinz-Wilhelm Hübers} received the Diplom and doctoral degree in physics from the Universität Bonn, Bonn, Germany, in 1991 and 1994, respectively. From 1991 to 1994, he was with the Max-Planck-Institut für Radioastronomie, Bonn, Germany. In 1994, he joined the Deutsches Zentrum für Luft- und Raumfahrt (German Aerospace Center, DLR), Berlin, Germany, becoming the Head of Department in 2001. From 2009 to 2014, he has been a Professor of Experimental Physics with the Technische Universität Berlin, Berlin, Germany, and the Head of the Department “Experimental Planetary Physics” at DLR. In 2014, he became the Director of the Institute of Optical Sensor Systems, DLR, and a Professor with the Humboldt-Universität zu Berlin. His research interests include THz physics and spectroscopy, particularly in THz systems for astronomy, planetary research, and security. Prof. Hübers has received the Innovation Award on Synchrotron Radiation (2003) and the Lilienthal Award (2007). In 2021 he will receive an honorary doctorate at Chalmers University of Technology, Gothenburg, Sweden.
\end{IEEEbiography}

\begin{IEEEbiography}[{\includegraphics[width=1in,height=1.5in,clip,keepaspectratio]{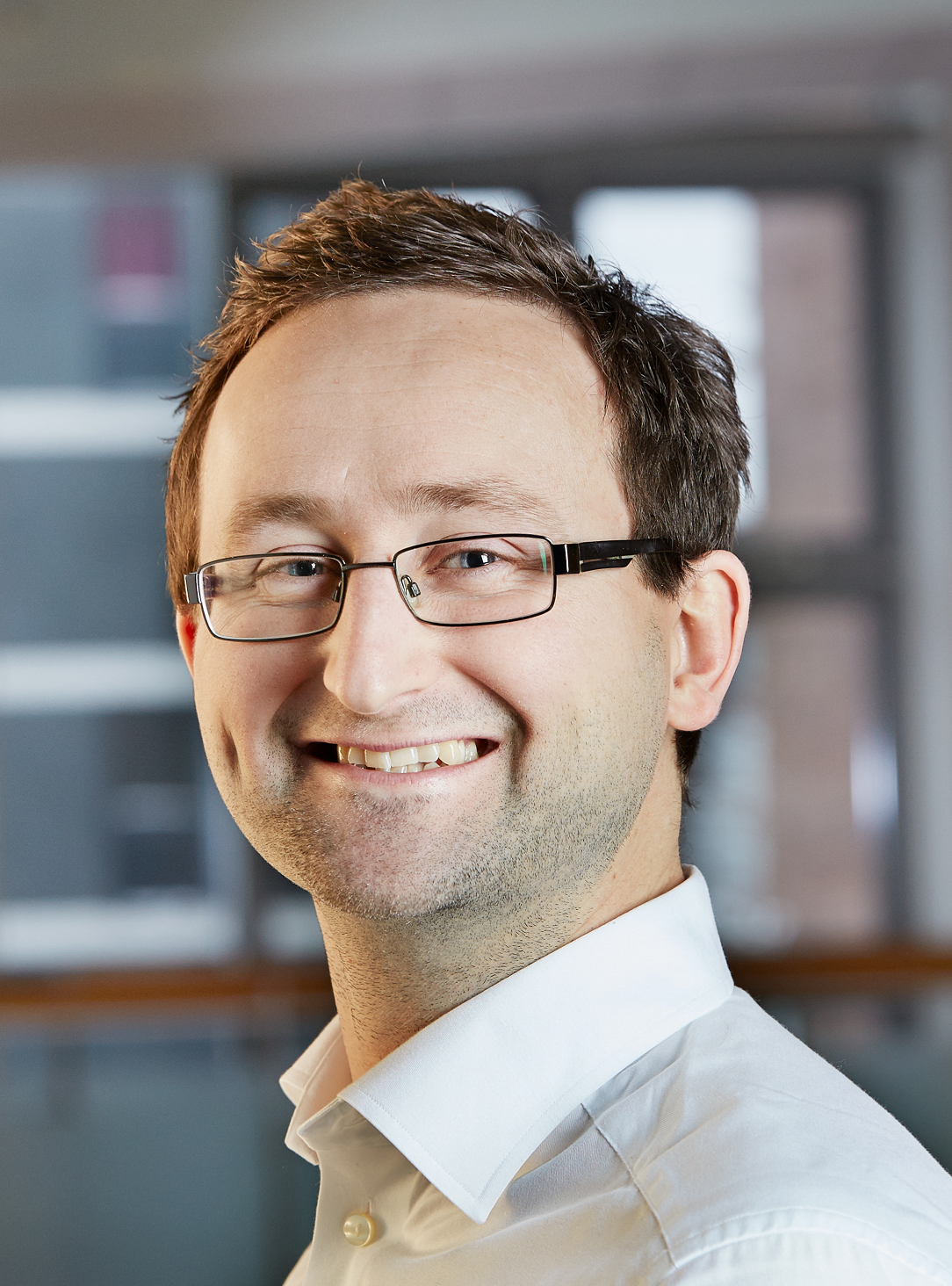}}]%
{Peter Sobis}
(S’05) was born in Gothenburg, Sweden, in 1978. He received the M.Sc. degree in electrical engineering in 2003 and the Licentiate and Doctoral degrees in THz electronics in 2010 and 2016, respectively, from the Chalmers University of Technology, Gothenburg, Sweden.
From 2003 to 2004, he was with Anaren Microwave Inc. in Syracuse, NY, USA, working on passive microwave components and beamforming networks. In 2004, he joined Omnisys Instruments AB, Västra Frölunda, Sweden, where he has been responsible for the development of radiometer components and subsystems for various ESA and Swedish National Space Board projects. In 2018, he became Adjunct Professor at the Department of Microtechnology and Nanoscience (MC2) at Chalmers University of Technology. His current research involves Schottky-based receiver systems, measurement techniques, and the development of multi-functional radiometer chipsets and integrated modules for earth observation and radio astronomy applications.
\end{IEEEbiography}

\begin{IEEEbiography}[{\includegraphics[width=1in,height=1.5in,clip,keepaspectratio]{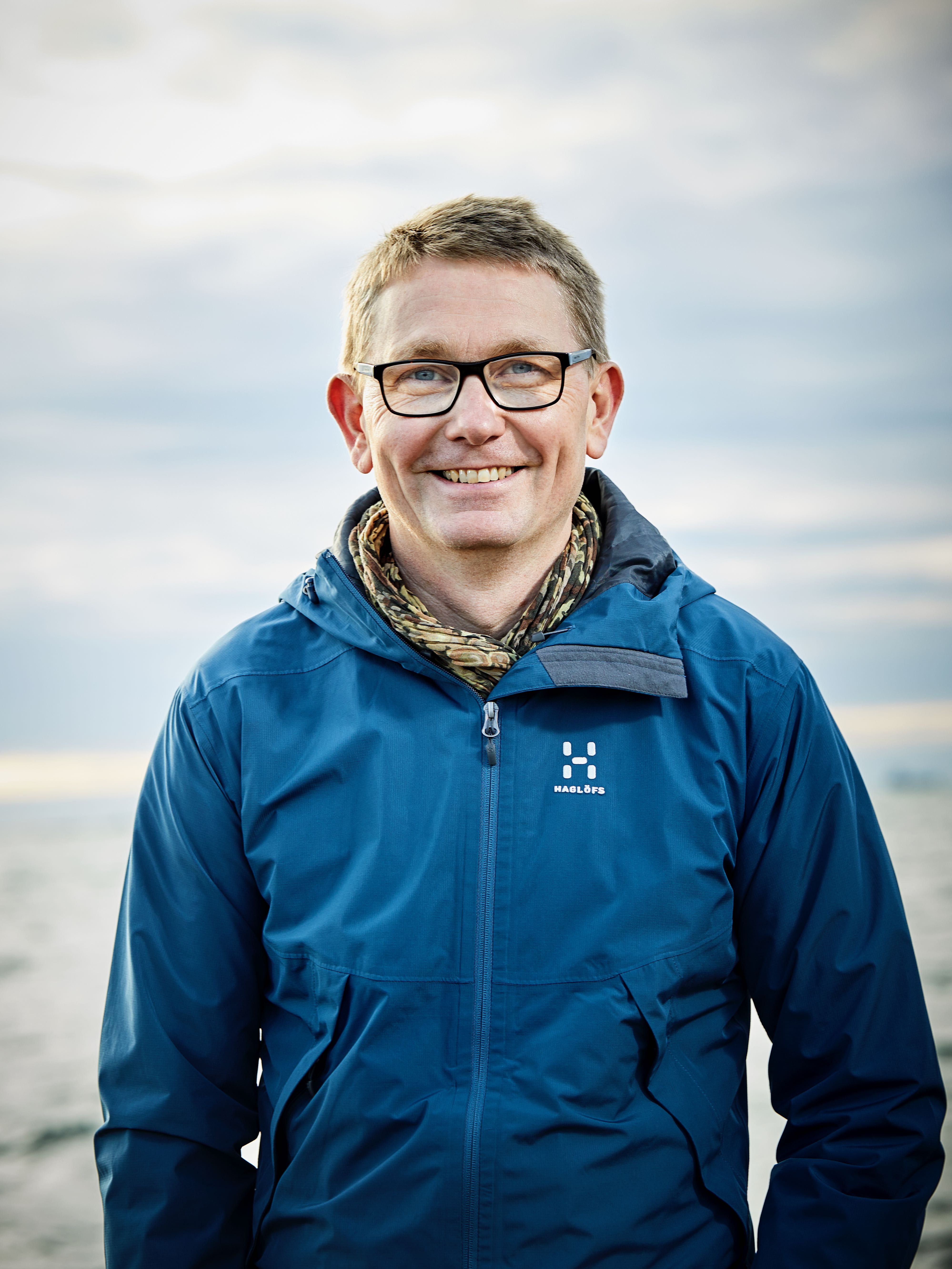}}]%
{Jan Stake}
(S’95–M’00–SM’06) was born in Uddevalla, Sweden, in 1971. He received the M.Sc. degree in electrical engineering and the Ph.D. degree in microwave electronics from the Chalmers University of Technology, Gothenburg, Sweden, in 1994 and 1999, respectively.

In 1997, he was a Research Assistant with the University of Virginia, Charlottesville, VA, USA. From 1999 to 2001, he was a Research Fellow with the Millimetre Wave Group at the Rutherford Appleton Laboratory, Didcot, UK. He then joined Saab Combitech Systems AB, Linköping, Sweden, as a Senior RF/microwave Engineer, until 2003. From 2000 to 2006, he held different academic positions with the Chalmers University of Technology and from 2003 to 2006, he was also the Head of the Nanofabrication Laboratory, Department of Microtechnology and Nanoscience (MC2). During 2007, he was a Visiting Professor with the Sub-millimetre Wave Advanced Technology (SWAT) Group at Caltech/JPL, Pasadena, CA, USA. In 2020, he was a Visiting Professor at TU Delft. He is currently a Professor and the Head of the Terahertz and Millimetre Wave Laboratory, Chalmers University of Technology. He is also the Co-founder of Wasa Millimeter Wave AB, Gothenburg, Sweden. His research interests include graphene electronics, high frequency semiconductor devices, THz electronics, sub-millimetre wave measurement techniques (THz metrology), and THz in biology and medicine.
\newline
Prof. Stake served as the Editor-in-Chief for the IEEE Transactions on Terahertz Science and Technology between 2016 and 2018 and Topical Editor between 2012 and 2015.
\end{IEEEbiography}

\end{document}